\documentclass[journal=jctcce,manuscript=article]{achemso}
\usepackage[version=3]{mhchem}
\usepackage{caption}
\usepackage{subcaption}
\usepackage{hyperref}
\usepackage{natbib}
\usepackage{siunitx}
\usepackage{tabularx}
\usepackage{amsmath}
\usepackage{amssymb}
\usepackage{xcolor}
\usepackage{threeparttable}
\usepackage{multirow}
\usepackage{booktabs}
\usepackage{subcaption}
\usepackage{braket}
\usepackage{derivative}
\usepackage{physics}
\usepackage{gensymb}
\newcommand{\bfR}{{\mathbf{R}}}
\newcommand{\bfD}{{\mathbf{D}}}
\newcommand{\bfU}{{\mathbf{U}}}
\newcommand{\bfX}{{\mathbf{X}}}
\newcommand{\bfA}{{\mathbf{A}}}
\newcommand{\bfM}{{\mathbf{M}}}
\newcommand{\bfr}{{\mathbf{r}}}

\newcommand{\bfs}{{\mathbf{s}}}
\newcommand{\bfzero}{{\mathbf{0}}}
\newcommand{\bfOmega}{{\mathbf{\Omega}}}
\newcommand{\bftau}{{\boldsymbol{\tau}}}
\newcommand{\h}{\hat{\mathbf{\mathcal{H}}}}
\newcommand{\ha}{\text{H}}
\newcommand{\qa}{\text{Q}}
\newcommand{\V}{\mathcal{V}}
\newcommand{\aobs}{\mathcal{O}}

\author{Arpan Kundu}
\affiliation{Pritzker School of Molecular Engineering, The University of Chicago, Chicago, Illinois 60637, United States}
\email{arpank@uchicago.edu, arpan.kundu@gmail.com}
\author{Giulia Galli}
\affiliation{Department of Chemistry, University of Chicago, Chicago, Illinois 60637, United States}
\alsoaffiliation{Pritzker School of Molecular Engineering, The University of Chicago, Chicago, Illinois 60637, United States}
\alsoaffiliation{Materials Science Division and Center for Molecular Engineering, Argonne National Laboratory, Lemont, Illinois 60439, United States}
\email{gagalli@uchicago.edu}

\title{Quantum vibronic effects on the electronic properties of molecular crystals}
\begin{document}
\maketitle

\begin{abstract}
We present a study of molecular crystals, focused on the effect of nuclear quantum motion and anharmonicity on their electronic properties. We consider a system composed of relatively rigid molecules, a diamondoid crystal, and one composed of floppier molecules, NAI-DMAC, a thermally activated delayed fluorescence compound. We compute fundamental electronic gaps at the DFT level of theory, with the PBE and SCAN functionals, by coupling first-principles molecular dynamics with a nuclear quantum thermostat. We find a sizable zero-point-renormalization (ZPR) of the band gaps, which is much larger in the case of diamondoids (~ 0.6 eV) than for NAI-DMAC (~ 0.22 eV). We show that the frozen phonon (FP) approximation, which neglects inter-molecular anharmonic effects,  leads to a large error $(\sim 50\%)$ in the calculation of the band gap ZPR. Instead, when using a stochastic method, we obtain results in good agreement with those of our quantum simulations for the diamondoid crystal. However, the agreement is worse for NAI-DMAC where intra-molecular anharmonicities contribute to the ZPR.  Our results highlight the importance of accurately including nuclear and anharmonic quantum effects to predict the electronic properties of molecular crystals. 
\end{abstract}

\section{Introduction}\label{sec:intro}

 Quantum vibronic effects play an important role in determining the temperature dependence of the electronic properties of molecules and solids, including their fundamental electronic gaps, as reported for small molecules\cite{Shang_JPCA_2021,Han_JCTC_2022}, nano-clusters\cite{FPH_Bester_2013,FPH_Bester_2016,Ryan_JCTC_2018, Kundu_PRM_2021}, crystalline \cite{Kundu_PRM_2021, Giustino_PRL_2010, Antonius_PRL_2014, Monserrat_PRB_2014, Karsai_NJP_2018, Han_JCTC_2021, Monserrat_PRB_2015_mol_cry, Altvater_PRB_2020, Alvertis_PRB_2022}, and amorphous solids \cite{Kundu_PNAS_2022}. 
Hence predicting quantum vibronic effects is critical to understanding the physical behavior of systems used in applications ranging from bio-electronics, thermo-electrics and photovoltaics, to optical fiber technologies, spintronics and quantum sensing.

In first principle electronic structure calculations of electron-phonon interactions,  quantum vibronic effects are  included using perturbative\cite{Giustino_Rev_2017} or non-perturbative methods \cite{Monserrat_Rev_2018} such as frozen phonon (FP) and stochastic methods. All of these techniques rely on the harmonic approximation of the potential energy surface (PES). In addition, the FP method assumes that electronic eigenenergies are quadratic with respect to given phonon modes' coordinates, and  perturbative methods also invoke the rigid ion approximation to represent the nuclear Hamiltonian. Previous work showed that the rigid ion approximation is not adequate to compute the renormalization of the fundamental gap of isolated molecules \cite{Kundu_PRM_2021, Gonze_2011},  and that the quadratic approximation is not sufficiently accurate for molecular crystals composed of small molecules \cite{Monserrat_PRB_2015_mol_cry}. Furthermore, recently  the validity of the harmonic approximation has been  questioned  in the case of perovskites \cite{Carbogno_PRB_2020}, organic molecular crystals \cite{Alvertis_PRB_2022}, and amorphous carbon\cite{Kundu_PNAS_2022}.  

The stochastic method is an interesting alternative to the FP technique, especially the so-called one-shot implementation \cite{Zacharias_PRL_2015, Zacharias_PRB_2016} proposed by Zacharias et al. 
Using such method, Monserrat et al reported very large (1-2 eV) zero phonon renormalizations (ZPR) of the fundamental gap of molecular crystals composed of small molecules: CH\textsubscript{4}, NH\textsubscript{3}, H\textsubscript{2}O and HF,  and discussed the limitations of the quadratic approximation within the FP method \cite{Monserrat_PRB_2015_mol_cry}. Using path-integral molecular dynamics simulations and a machine-learned potential, Alvertis et al showed that the harmonic approximation, which underlies both the stochastic and FP methods,  fails catastrophically for acene molecular crystals \cite{Alvertis_PRB_2022} and their results challenged the validity of the stochastic method, at least for some molecular crystals.   However, the authors of Ref.\cite{Alvertis_PRB_2022} also pointed out that the short-range nature of the machine-learned potentials employed in their study may introduce non-negligible residual error in band gap renormalizations.       

 First-principles molecular dynamics (FPMD) simulations \cite{Marx_Hutter_book} can fully account  for anharmonic vibronic effects, but only at the classical level, yielding reliable results   near or above the Debye temperature. Path-integral FPMD simulations \cite{PI_Berne_Rev_1986, PI_Marx_Rev_1996, PI_hererro_rev_2014} constitute an accurate framework to describe nuclear quantum effects (NQEs), however, due to their  computational cost  they are  rarely adopted to study electron-phonon interactions in molecules and solids. Recently, we showed that using a colored noise generalized Langevin equation thermostat (also known as quantum thermostat\cite{QT_Ceriotti_PRL_2009, QT_Ceriotti_JCTC_2010, QT_Review_Finocchi_2022}), in conjunction with FPMD simulations, one can accurately predict the impact of NQEs on the electronic properties of several carbon-based systems with an affordable computational cost\cite{Kundu_PRM_2021}. 

Here we use FPMD simulations with a quantum thermostat to accurately include anharmonic quantum vibronic effects and study their impact on the electronic properties of molecular crystals. We then compare FPMD results  with those obtained using the FP and the stochastic one-shot methods to assess the validity of the approximations adopted when using these techniques. We investigate in detail two  molecular crystals: one composed of a rigid diamondoid molecule, [1(2,3)4]pentamantane\cite{Dahl_Science_2003}, and one composed of a floppy molecule, NAI-DMAC\cite{Zeng_Adv_Mater}, which has important applications for third-generation organic light-emitting diodes (OLED)\cite{Zeng_Adv_Mater}. We found that for both crystals, the frozen phonon method performs poorly. For the pentamantane crystal, the stochastic method yields much improved results compared to those of  FP. In contrast, for NAI-DMAC, the stochastic method does not lead to any substantial  improvement. To understand the origin of the breakdown of the different approximations, we also perform calculations for isolated molecules which shed light on the importance of intra- and inter-molecular anharmonicities on the vibronic coupling, and hence on their impact on electronic properties.

The rest of the paper is organized as follows. In section \ref{sec:method} we describe the methods adopted in our work. We present our results for the pentamantane molecule and molecular crystal in section \ref{sec:validation} and \ref{sec:penta-crystal}, respectively. The results obtained for both the isolated molecule and the crystal of NAIDMAC are discussed in section \ref{sec:naidmac}. Finally, we give our summary and conlusions in section \ref{sec:conclusion}.     

\section{Methods}\label{sec:method}
\subsection{Theory}
Within the Born-Oppenheimer (BO) approximation, the Schr\"odinger equation of a system with $N$ nuclei can be expressed as\cite{Monserrat_Rev_2018},
\begin{equation} \label{eq:vib_se}
\h |\chi_k(\bfR) \rangle = \qty(-\frac{1}{2}\sum_{I}\frac{1}{M_I}\nabla_{R_I}^2 + \V(\bfR)) |\chi_k(\bfR) \rangle = \varepsilon_k |\chi_k(\bfR) \rangle,
\end{equation}
where $M_I$ and $\bfR = (R_{1x},R_{1y},...,R_{Nz})$ are the mass of the $I$-th nucleus and the Cartesian position vector of all nuclei with respect to a chosen origin, respectively. Here $\bfr$  and $|\psi(\bfr; \bfR) \rangle$ are the electronic coordinates and wave function, respectively. $\V(\bfR)$, and $|\chi_k(\bfR)\rangle$ represent the $3N$-dimensional adiabatic potential energy surface, and the wave function for the k-th nuclear state, respectively. 
%Since, $\V(\bfR)$ and $|\chi_k(\bfR)\rangle$ are indifferent to the choice of the origin, for the convenience of discussion, we set the origin at the coordinates of the equilibrium geometry where $\V(\bfR)$ is minimum. 
For simplicity, we consider only the vibrational modes at the $\Gamma$-point of the simulation supercell. 

%We consider  electronic eigenenergies or total energy, $\aobs(\bfR) = \langle \psi(\bfr; \bfR) | \hat{\aobs}(\bfr; \bfR) | \psi(\bfr; \bfR) \rangle $, that depends on one electronic state. 
%Here $\bfr$  and $|\psi(\bfr; \bfR) \rangle$ are the electronic coordinates and wave function, respectively. 
%We note that as a consequence of the Born-Oppenheimer (BO) approximation, the electronic wave function has only parametric dependence on the nuclear coordinates, $\bfR$. 
When the system is at equilibrium at a temperature $T$,  the effect of electron-phonon interaction on the electronic property $\aobs(\bfR)$ can be included by performing an ensemble average over all adiabatic nuclear states $k$,
\begin{equation} \label{eq:ensemble_avg}
\begin{aligned}
\langle \aobs \rangle_T = \frac{1}{Q(T)}\sum_{k=0}^{\infty} \langle\chi_k(\bfR) | \hat{\aobs}(\bfR) |\chi_k(\bfR) \rangle \exp(-\frac{\varepsilon_k}{k_BT}) = \int d\bfR W(\bfR,T) \aobs (R).
\end{aligned}
\end{equation}
Here,
\begin{equation}
    W(\bfR,T) = \frac{1}{Q(T)} \sum_{k=0}^{\infty}\langle \chi_k(\bfR) | \chi_k(\bfR) \rangle \exp(-\frac{\varepsilon_k}{k_BT}),
\end{equation}
denotes the probability of finding the system with the nuclear coordinates within $\bfR$ and $\bfR+d\bfR$.
The partition function $Q(T)$ is defined as, $Q(T) = \sum_{k=0}^{\infty} \exp(-\frac{\varepsilon_k}{k_BT})$, where $k_B$ is the Boltzmann constant. 

A molecular dynamics simulation with either a path-integral approach\cite{PI_Berne_Rev_1986,PI_hererro_rev_2014} or quantum thermostat approach \cite{QT_Ceriotti_PRL_2009,QT_Ceriotti_JCTC_2010,QT_Review_Finocchi_2022} utilizes Eq. \ref{eq:ensemble_avg} to compute the electron-phonon renormalized electronic properties. However, being computationally expensive, such simulations are not often adopted for computing electron-phonon renormalizations of electronic properties from first principles. Commonly used approaches in solid-state physics employ the harmonic approximation (HA) to $\V(\bfR)$, 
%to simplify the expression in Eq. \ref{eq:ensemble_avg}. A Taylor series expansion of $\V(\bfR)$ near the equilibrium geometry, $\bfR = \bfzero$, and subsequently, neglecting the terms beyond second order yields the potential energy with harmonic approximation (HA),
\begin{equation} \label{eq:ha}
\V^{\ha}(\bfR) = \frac{1}{2}\sum_{I\alpha,J\alpha'}(R_{I\alpha}\sqrt{M_I})D_{I\alpha,J\alpha'}(\sqrt{M_J}R_{J\alpha'}) = \frac{1}{2}\bfR^{T}\bfM^{1/2}\bfD\bfM^{1/2}\bfR,
\end{equation}
with $\bfM$ representing a $3N\times3N$ diagonal matrix of nuclear masses. 
%We note that the first order term of the Taylor expansion becomes zero by setting the condition of the minimum, $\pdv{\V(\bfR)}{R_{I\alpha}}=0$. In addition, we set the reference potential energy, $\V(\bfzero)=0$. 
The elements of the dynamical matrix, also known as the mass-weighted Hessian matrix, are given by,
\begin{equation} \label{eq:dynmat}
      D_{I\alpha,J\alpha'} = \frac{1}{\sqrt{M_{I}M_{J}}}\eval{\pdv{\V(\bfR)}{R_{I\alpha}}{ R_{J\alpha'}}}_{\bfR=\bfzero},
\end{equation}
where $\alpha$,$\alpha'$ denote the cartesian axes $x$,$y$ or $z$ and $I,J$ denote the indices of the nuclei. 
%Spectral decomposition of the dynamical matrix, $\bfD = \bfU\bfOmega^2\bfU^T$, returns a unitary matrix, $\bfU$, and a $3N\times3N$ diagonal matrix of normal-mode frequencies, $\bfOmega$, with diagonal elements: $\omega_1,\omega_2,...,\omega_{3N}$. The unitary matrix, $\bfU$, defines the cartesian to normal mode transformations,
%\begin{equation}\label{eq:cart2norm}
%\begin{aligned}
%    \bfX=\bfU^T\bfM^{1/2}\bfR, \\
%     \nabla_X^2=\bfU^T\bfM^{1/2}\nabla_R^2,
%\end{aligned}
%\end{equation}
%and back transformation to cartesian from normal modes,
%\begin{equation}\label{eq:norm2cart}
%\begin{aligned}
%    \bfR=(\bfM^{1/2})^{-1}\bfU\bfX, \\
%     \nabla_R^2=(\bfM^{1/2})^{-1}\bfU\nabla_X^2,
%\end{aligned}
%\end{equation}
%with $\bfX$ representing the matrix of $3N$ normal mode vectors.

A spectral decomposition of the dynamical matrix, $\bfD = \bfU\bfOmega^2\bfU^T$, returns a unitary matrix, $\bfU$, and a $3N\times3N$ diagonal matrix of normal-mode frequencies, $\bfOmega$, with diagonal elements: $\omega_1,\omega_2,...,\omega_{3N}$. The total nuclear Hamiltonian can be separated into 3 independent translational degrees of freedom, $d_r$ number of independent global rotational degrees of freedom, and $3N-3-d_r$ number of independent vibrational degrees of freedom. 
For solid, linear, and non-linear isolated molecules, the number of rotational degrees of freedom ($d_r$) is 0, 2, and 3, respectively. 
Since the translations and global rotations, which appear as the first $3+d_r$ lowest eigenvalues of $\Omega$, do not affect the electronic properties, we focus on the vibrational Hamiltonian: 
\begin{equation}
    \h^{\ha}  = \sum_{\nu=3+d_r+1}^{3N}\qty(-\frac{1}{2}\nabla_{X_\nu}^2+\frac{1}{2}\omega_\nu^2X_\nu^2).
\end{equation}
with $X_\nu$ denoting the $\nu$-th normal mode.
%The wavefunctions, $|\chi_{\nu,k}^\ha \rangle$, and energies, $\varepsilon_{\nu,k}^\ha$ of each simple harmonic oscillator is known analytically, 
%\begin{equation}\label{eq:ha_wf}
%\langle X_\nu | \chi_{\nu,k}^\ha \rangle =  \frac{1}{\sqrt{2^kk!}}\qty(\frac{\omega_\nu}{\pi})^\frac{1}{4}\exp\qty[\frac{-\omega_\nu X_\nu^2}{2}]H_k\qty(\sqrt{\omega_\nu}X_\nu),
%\end{equation}
%\begin{equation}\label{eq:ha_en}
%    \varepsilon_{\nu,k}^\ha = \qty(k+\frac{1}{2})\omega_\nu,
%\end{equation}
%with $H_k$ denoting the $k$-th order Hermite polynomial. Inserting the expression of $\varepsilon_{\nu,k}^\ha$ from Eq. \ref{eq:ha_en} into Eq. \ref{eq:partion_fn} yields, the partition function for the $\nu$-th harmonic oscillator,
The partition function for the $\nu$-th harmonic oscillator is
\begin{equation}\label{eq:ha_pf}
   Q_{\nu}^\ha(T) = \sum_{k=0}^{\infty} \exp\qty[-\frac{\omega_\nu}{k_BT}(k+\frac{1}{2})] = \exp\qty(\frac{\omega_\nu}{2k_BT})n_B(\omega_\nu,T),
\end{equation}
where the Bose occupation factor is given by,
\begin{equation}\label{eq:be_occ}
    n_B(\omega,T) = \frac{1}{\exp(\omega/k_BT)-1}
\end{equation}
%Because of the separation of Hamiltonian into $3N-3-d_r$ number of independent vibrational terms, the total vibrational wave function which is characterized by a vector of quantum numbers $\bfk = \qty(k_{3+d_r+1},k_{3+d_r+2},...,k_{3N})$ can be simplified by the product of the wave functions for each individual normal mode. The total partition function can be written analogously.
%\begin{equation}\label{eq:ha_tot_wf_pf}
%    \begin{aligned}
%        |\chi_{\bfk}^\ha(R)\rangle = \prod_{\nu=3+d_r+1}^{3N}|\chi_{\nu,k_\nu}^\ha(R)\rangle, \\
%        Q^\ha(T)=\prod_{\nu=3+d_r+1}^{3N}Q_\nu^\ha(T).
%    \end{aligned}
%\end{equation}
%Utilizing the expressions from Eqs.\ref{eq:ha_wf}--\ref{eq:ha_tot_wf_pf}, Eq. \ref{eq:ensemble_avg} can be rewritten as,
%Utilizing thee expressions from \ref{eq:ha_tot_wf_pf}, Eq. \ref{eq:ensemble_avg} can be rewritten as,
As a consequence of the separable form of the Hamiltonian, Eq. \ref{eq:ensemble_avg} can be simplified:
\begin{equation}\label{eq:ha_avg}
    \langle \aobs \rangle_T^\ha = \int d\bfX W^\ha(\bfX,T) \aobs(\bfX) = \int d\bfX \qty[\prod_{\nu=3+d_r+1}^{3N} G(X_\nu;\sigma_{\nu,T})]\aobs(\bfX) 
\end{equation}
where the harmonic probability density, $W^\ha(\bfX,T)$, reduces to a product of independent Gaussian functions, $G(X_\nu;\sigma_{\nu,T}) = \frac{1}{\sqrt{2\pi\sigma_{\nu,T}^2}}\exp\qty(-\frac{X_{\nu}^2}{2\sigma_{\nu,T}^2})$,
with widths related to the Bose occupation factor:
\begin{equation}\label{eq:sigma}
    \sigma_{\nu,T}= \sqrt{\frac{2n_B\qty(\omega_\nu,T)+1}{2\omega_\nu}}.
\end{equation}

We note that though Eq. \ref{eq:ha_avg} is valid under the harmonic approximation, it does not assume any explicit dependence of the electronic observable $\aobs$ on nuclear coordinates ($\bfR$) or normal mode coordinates ($\bfX$). To further simplify the expression, we Taylor expand  $\aobs(\bfX)$:
\begin{equation}\label{eq:obs_taylor}
    \aobs(\bfX) = \aobs(\bfzero) + \sum_{\nu}\eval{\pdv{\aobs}{X_\nu}}_{\bfzero}X_{\nu} + \sum_{\nu\nu'} \eval{\pdv{\aobs}{X_{\nu}}{X_{\nu'}}}_{\bfzero}X_{\nu}X_{\nu'} + ...
\end{equation}
and truncate the expansion after the second order. After inserting the resulting expression into Eq.\ref{eq:ha_avg}, we obtain the phonon renormalized electronic observable within the quadratic (Q) approximation,
\begin{equation}\label{eq:qa_avg}
    \langle \aobs \rangle_T^\qa = \aobs(\bfzero) +  \sum_{\nu=3+d_r+1}^{3N}\frac{1}{2\omega_\nu}\eval{\pdv[2]{\aobs}{X_{\nu}}}_0\qty[n_B(\omega_{\nu},T)+\frac{1}{2}].
\end{equation}
We note that terms of odd order in $X_\nu$ or cross-coupling terms such as $X_{\nu}X_{\nu'}$, with $\nu \neq \nu'$, do not appear because in the harmonic approximation the density is symmetric with respect to $\bfX = \bfzero$. For systems with strong anharmonicity, the vibrational density would no longer be symmetric, and hence both odd-order terms and cross-coupling terms  would become important.

\subsection{Stochastic Approach}
The stochastic approach employs Monte Carlo sampling to evaluate $W(\bfX,T)$ and  Eq. \ref{eq:ha_avg} to compute phonon-renormalized electronic properties. At each Monte Carlo step, a displaced normal mode coordinate is obtained, $\bfX^i=\bftau^{i}$, where, for $\nu>3+d_r$, the matrix elements, $\tau_{\nu}^i$, are a Gaussian distributed random number with zero mean and width $\sigma_{\nu,T}$, while the first $3+d_r$ matrix elements are set to zero. Then, the $\bfX^i$'s are back-transformed to Cartesian coordinates, and $\bfR^i$, and the electronic observable $\aobs(\bfR^i)$ are computed. After $M$ Monte Carlo steps, Eq. \ref{eq:ha_avg} can be re-written as,
\begin{equation}\label{eq:mc_avg}
        \langle \aobs \rangle_T^{MC} = \frac{1}{M}\sum_{i=1}^{M}\aobs(\bfX^i) = \frac{1}{M}\sum_{i=1}^{M}\aobs(\bfR^i)
\end{equation}

Based on the Mean-value (MV) theorem and utilizing a quadratic (Q) approximation, Monserrat showed that there exists $2^{3N-3-d_r}$ mean-value positions, $\bfX_\text{MVQ}$, for which $\aobs(\bfX_\text{MVQ})\simeq\langle \aobs \rangle_T$,\cite{TL_Monserrat_PRB_2016} with $\bfX^i_\text{MVQ} =  \bfs^i\boldsymbol{\sigma}_T$, where the matrix elements of $\boldsymbol{\sigma}_T$ is given by Eq. \ref{eq:sigma}, and $\bfs^i$ is a matrix with the first $3+d_r$ elements set to zero and the remaining $3N-3-d_r$ elements being either +1 or -1. It was shown that a Monte Carlo algorithm that samples random signs, $s^i$, has a faster convergence for the value of $\langle \aobs \rangle_T^{MC}$ than the one that samples  random numbers, $\boldsymbol{\tau}^i$, from a Gaussian distribution. \cite{TL_Monserrat_PRB_2016}. Following Monserrat's work,  Zacharias and Giustino proposed a one-shot (OS) algorithm, \cite{Zacharias_PRB_2016} in which the signs are chosen according to:
\begin{equation}
\begin{split}
    s_{\nu} &= (-1)^{\nu-4-d_r} \text{ for } \nu > 3+d_r \\
    &= 0, \text{ for } \nu \le 3+d_r
\end{split}
\end{equation}
and they showed that only a single first-principle calculation for the atomic configuration, $\bfX=\bfs\boldsymbol{\sigma}_T$ is sufficient to converge the value of $\langle \aobs \rangle_T$. They also proposed that an additional first-principles calculation on the antithetic pair of the chosen atomic configuration, i.e., $\bfX=-\bfs\boldsymbol{\sigma}_T$, can improve the result, and this is the approximation  adopted here. 
\begin{equation}\label{eq:mc_avg}
        \langle \aobs \rangle_T^{OS} = \frac{1}{2}\qty[\aobs\qty(\bfs\boldsymbol{\sigma}_T) + \aobs\qty(-\bfs\boldsymbol{\sigma}_T)]
\end{equation}

\subsection{Frozen Phonon Approach}
A frozen phonon (FP) approach utilizes Eq. \ref{eq:qa_avg} to compute phonon-renormalized electronic properties. Throughout this work, our electronic observables $(\aobs)$ are either the valence and conduction band (VB) energies $(E_{n})$ or the band gap $(E_g)$. The second derivative of the $n$-th band energy with respect to $\nu$-th phonon mode scaled by the phonon frequency (see Eq. \ref{eq:qa_avg} when $\aobs=E_n$) is called the electron-phonon coupling energy (EPCE).
\begin{equation}\label{eq:epce}
    \text{EPCE}_{n,\nu} = \frac{1}{2\omega_\nu}\eval{\pdv[2]{E_n}{X_{\nu}}}_0
\end{equation}
It is evident from Eq. \ref{eq:qa_avg}, that the electron-phonon renormalization of the $n$-th band, 
\begin{equation}
    \Delta E_n(T) = \langle E_n \rangle_T - E_n(\bfzero)
\end{equation}
reduces to,
\begin{equation}
     \Delta E_n^Q(0) = \sum_{\nu=3+d_r+1}^{N}\text{EPCE}_{n,\nu}
\end{equation}
within the quadratic approximation at 0 K. The EPCEs can be computed using the FP method (See section S1.3 in the SI for more details).

%The value of $h$ is chosen such that the harmonic approximation predicts a 0.001 a.u. energy displacement along each mode, i.e., $0.5\times\omega_{\nu}^2\times h = 0.001$. Using these displaced coordinates, the first and second derivatives of the n-th band energy, $E_{n}$ is 

\subsection{Computational Details}

We used the Qbox code \cite{Qbox_Gygi_2008} for the optimization of the geometry and cell parameters of all systems studied here. For molecular dynamics (MD) simulations with a quantum thermostat(QT), which in the following we call "QTMD" simulations, we used Qbox coupled\cite{Kundu_PRM_2021} to the i-PI code, where the i-PI driver \cite{ipi_Kapil_2018} moves the nuclear coordinates and Qbox compute forces from density functional theory (DFT). To obtain the zero-point renormalization (ZPR) of electronic energies, $\Delta E_{n}(0)$, the electron-phonon renormalizations at finite $T$, $\Delta E_{n}(T)$ are fitted with the Vi\~{n}a model \cite{Vina_PRB_1984}:
\begin{equation} \label{eq:vina_eq}
    \Delta E_{n}(T)=\Delta E_{n}(0)-A[1+2/(e^{B/T}-1)]
\end{equation}

For the frozen phonon and stochastic calculations, we used the PyEPFD package\cite{pyepfd} to generate displaced structures and Qbox for DFT calculations. Specifically, the Qbox outputs were post-processed with the PyEPFD package to compute the dynamical matrix, phonon frequencies, phonon eigenvectors, and renormalized energy gaps. Throughout this work, we used a Cartesian displacement of 0.005 a.u. to compute the dynamical matrix elements. To compute the first and second derivatives, $O'$ and $O''$  we used a normal mode displacement that corresponds to a potential energy change, $\delta\V^{H} = 0.001$ a.u, see Eqs. S23-S25 in the SI.  In addition, we used PyEPFD to calculate the anharmonic measure\cite{Knupp_anh_mes_PRM_2020}  and vibrational densities along a phonon mode by post-processing quantum molecular dynamics trajectories obtained from QTMD simulations. 

For the pentamantane molecule, we used a generalized gradient approximated (GGA) Perdew-Burke-Erzenhof (PBE) functional \cite{PBE_Perdew_PRL_1996_1,PBE_Perdew_PRL_1996_2} and  strongly constrained and  approximately normed (SCAN)\cite{SCAN_PRL_2015} meta-GGA functional. For the molecular crystal of pentamantane, we only used the SCAN functional. For both the molecule and the crystal of NAI-DMAC, we used only the PBE functional in order to be consistent with our previous study.\cite{Francese_PCCP_2022} We used norm-conserving pseudopotentials \cite{ONCV_2015} with 50 and 60 Ry kinetic energy cutoffs for pentamantane and NAI-DMAC, respectively.  

\section{Results and Discussion}
\subsection{Electron-phonon renormalization of the electronic properties of the Pentamantane Molecule: validation of stochastic one-shot method for finite systems}\label{sec:validation}

 Using the PBE and SCAN functionals, we compare the results obtained with the stochastic one-shot method with those obtained with QTMD simulations and the FP method \cite{Kundu_PRM_2021}.
 
 The static HOMO-LUMO gap computed using the SCAN functional (5.71 eV) is 680 meV larger than that computed using PBE. The gap of pentamantane estimated from optical absorption measurements in the gas phase  (which naturally include electron-phonon interactions) is 5.81 eV at ambient temperatures\cite{Landt_PRL_2009}. The computed exciton binding energy of the molecule is about 2.5 eV \cite{FPH_Bester_2016} and hence the estimated fundamental gap is $\simeq 8.3$ eV.
%From our simulations, we compute fundamental band gaps that do not account for the exciton binding energy, and therefore, we expect a larger value compared to the experiment. 
At 250 K, QTMD simulations with the PBE and SCAN functionals underestimate the fundamental gap  and yield 4.58 eV and 5.23 eV, respectively. The QTMD simulations predict a ZPR value of -498 meV with the SCAN functional, 42 meV larger than the PBE value (-456 meV). To obtain the ZPR value from experiments, measurements should be performed at a wide range of temperatures and the data extrapolated to 0 K. Unfortunately such data are not available at present in the literature.

\begin{figure}[tbhp]
\centering
\includegraphics[width=8.4cm]{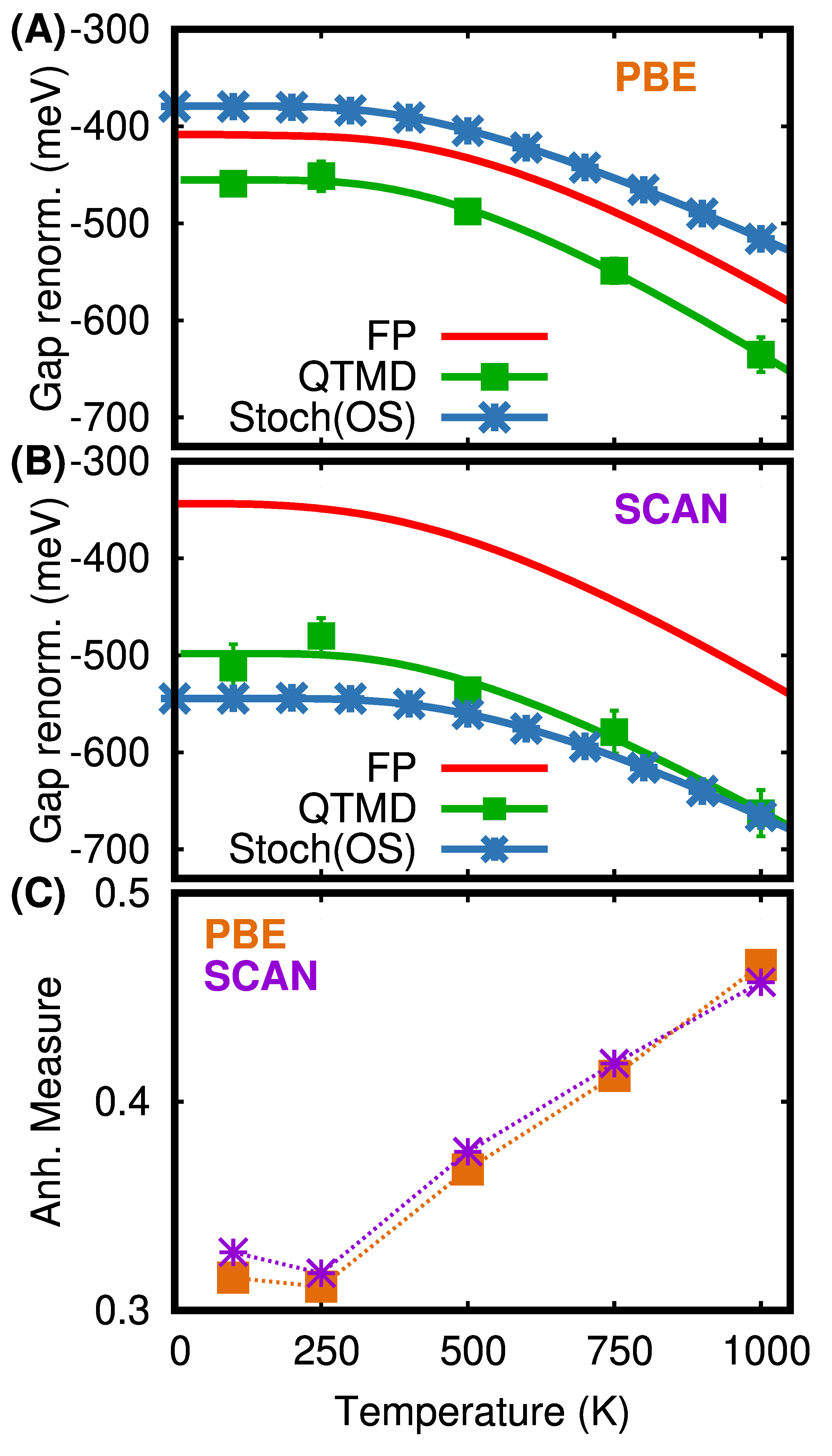}
\caption{Panels (A-B): Electron-phonon renormalization of the HOMO-LUMO gap of the pentamantane molecule as a function of temperature as computed using PBE (panel A) and SCAN (panel B) functionals. The gap renormalizations are computed using frozen phonon (FP), stochastic one-shot (Stoch(OS)), and first-principles molecular dynamics coupled with a quantum thermostat (QTMD). For Stoch(OS) and QTMD results, symbols represent the results obtained from simulations while the solid lines represent the Vi\~{n}a model fit of the simulation results. Panel C: Anharmonic measure for the pentamantane molecule at different temperatures computed from the trajectories obtained with QTMD simulations using the PBE and SCAN functionals. Broken lines are a guide to the eyes.}
\label{fig:penta-mol}
\end{figure}

 When using PBE (see Fig. \ref{fig:penta-mol}, panel A), the stochastic one-shot and FP results are in good agreement with each other, but  they deviate by 15\% from those of QTMD simulations, which fully account for anharmonic effects. 
%Since both FP and stochastic method relies on the harmonic vibrational density, the anharmonicity is the reason behind such deviation. Good agreement between FP and stochastic OS method suggests that the quadratic approximation for the electronic eigenvalues is a valid approach for an isolated pentamantane molecule when PBE functional is used. 

When using SCAN (see Fig \ref{fig:penta-mol}, panel B), we find that the FP method underestimates the gap renormalization by $\simeq 35\%$, while the stochastic one-shot method only slightly overestimates it. To investigate the reason for the overall good agreement of the stochastic and QTMD results for both the PBE and SCAN functionals and the worse agreement of FP and QTMD results in the case of SCAN, we computed an anharmonic measure proposed by Knupp et. al. \cite{Knupp_anh_mes_PRM_2020},
\begin{equation}\label{eq:anh_mes}
    a_x(T) =\sqrt{\frac{\sum_{x \in \bfA}\langle (F_x-F_x^\text{H})^2 \rangle_T}{\sum_{x \in \bfA}\langle F_x^2 \rangle_T}},
\end{equation}
for the configurations obtained from QTMD simulations. For a specific configuration, $F_x$ and $F_x^\text{H}$ denote the forces obtained for a coordinate $x$ from a first-principles calculation and from the harmonic approximation, respectively, and $\bfA$ denotes a subspace of Cartesian $(\bfR)$ coordinates for a set of specified nuclei or a subspace of normal-mode $(\bfX)$ coordinates. The total anharmonic measure is computed by including all atoms (or normal modes) in the subspace $\bfA$, while the anharmonic measure for a particular group of atoms or normal modes is computed by properly selecting the subspace $\bfA$. From Eq. \ref{eq:anh_mes}, it is easy to see that  anharmonic measure values larger than one indicate a breakdown of the harmonic approximation.   

Total anharmonic measures for the isolated pentamantane molecule as a function of T are shown in panel C of Fig. \ref{fig:penta-mol} (see also Fig. S1 in the SI for a normal-mode resolved anharmonic measures at 100K). We found similar values with the PBE and SCAN functional (below 0.5) up to  high temperatures, indicating that the harmonic approximation should perform relatively well and hence it is likely not the reason  of the worse performance of the FP method with the SCAN functional. Within the FP method, one employs the quadratic approximation for the HOMO-LUMO energies, which is not satisfied for the SCAN functional. Instead, the stochastic one-shot method does not rely on the quadratic approximation and includes electron-phonon couplings to any order, thus yielding results in  reasonable agreement with QTMD simulations for both functionals. The slight deviation (below $15\%$) originates from the harmonic approximation adopted in the stochastic method.

The stochastic one-shot method had been tested for several extended systems, but to the best of our knowledge, it has not yet been applied to finite systems.  Our results show that, as long as the harmonic approximation is valid, the one-shot method should also be applicable to finite systems and should perform better than the  FP method at a cheaper computational cost. Indeed, the FP method requires an order of $6N$ calculations to compute the second derivatives appearing in Eq. \ref{eq:qa_avg}, while the stochastic one-shot method requires only two calculations per temperature. Our results also indicate that  the validity of the quadratic approximation depends on the choice of the functional. The inclusion of intermediate range van-der Waals interactions in the meta-GGA scan functional may be a reason for a stronger non-quadratic electron-phonon coupling obtained with the SCAN functional.   

\subsection{Pentamantane molecular crystal}\label{sec:penta-crystal}
\begin{figure}[tbhp]
\centering
\includegraphics[width=16.4cm]{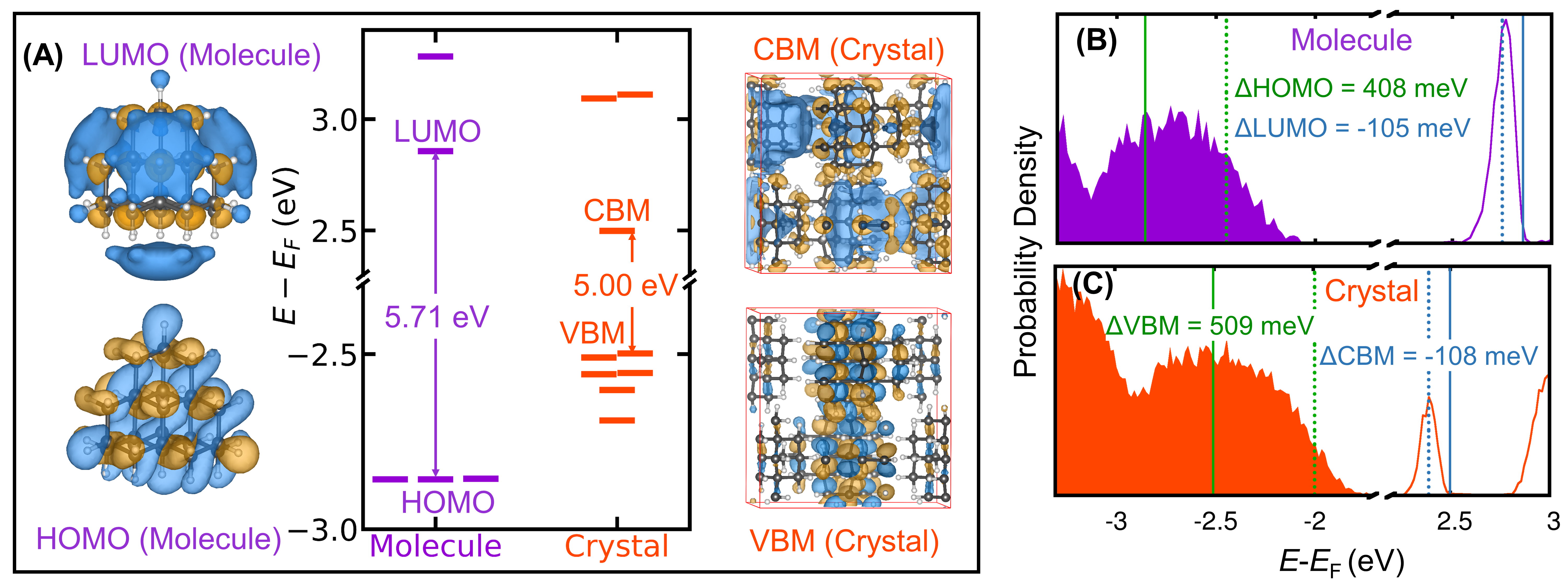}
\caption{Panel A shows the HOMO and LUMO levels relative to the Fermi level computed with the SCAN functional for an isolated pentamantane molecule and its molecular crystal when nuclei are at rest. Panels B and C show the electronic density of states at 100 K for the molecule and the crystal, respectively, computed using QTMD simulations with the SCAN functional. The green (blue) vertical lines represent the average energies of the HOMO or VBM (LUMO or CBM) when atoms are at rest (solid line) and at 100 K (broken line).}
\label{fig:homo-lumo-penta}
\end{figure}

 Now we turn to discuss the results for the molecular crystal of pentamantane. We used a unit cell containing 4 molecular units, with 232 atoms in total. Fig. \ref{fig:homo-lumo-penta}A, compares the HOMO (LUMO) of the molecule with the VBM (CBM) of the crystal unit cell when nuclei are at rest. In the molecule, the HOMO is three-fold degenerate, while the LUMO is singly degenerate. In the crystal, there are several valence band states close in energy to the VBM, the CBM remains singly degenerate. The static fundamental gap of the crystal is 0.71 eV smaller than that of the molecule when the SCAN functional is used. 
 We also computed the inverse participation ratios (IPR) \(\int |\psi|^4 \,d^3r\big /\big (\int |\psi|^2 \,d^3r\big)^2\) for the molecule's HOMO, LUMO and crystal's VBM, CBM which are $2.5\times10^{-4}$, $1.1\times10^{-5}$, $1.5\times10^{-4}$, and $1.7\times10^{-5}$, respectively, suggesting that the LUMO(CBM) is more delocalized. Fig \ref{fig:homo-lumo-penta}A also shows that the LUMO(CBM) is a surface state, with the CBM delocalized over two molecular units. In contrast, the HOMO(VBM) is localized between the C---C and C---H bonds, though the VBM is also spread out over two molecules.   
  
Panel B and C show the electronic density of states for the molecule and the crystal, respectively, obtained by including quantum nuclear vibrations at 100 K with QTMD simulations. In both cases, HOMO(VBM) and LUMO(CBM) move towards the Fermi level and consequently, the gap decreases when adding quantum vibronic effects. Moreover, the value of the HOMO(VBM) renormalization is much larger than that of the LUMO(CBM), due to its delocalized character. 
The renormalization of LUMO (CBM) of the molecule and the crystal are very similar, while the renormalization of the VBM of the crystal (-509 meV) is about 100 meV larger than HOMO of the molecule. The VBM of the crystal is localized on two molecular units and therefore, it is affected by both intra- and inter-molecular vibrations.  

\begin{figure}[tbhp]
\centering
\includegraphics[width=8.4cm]{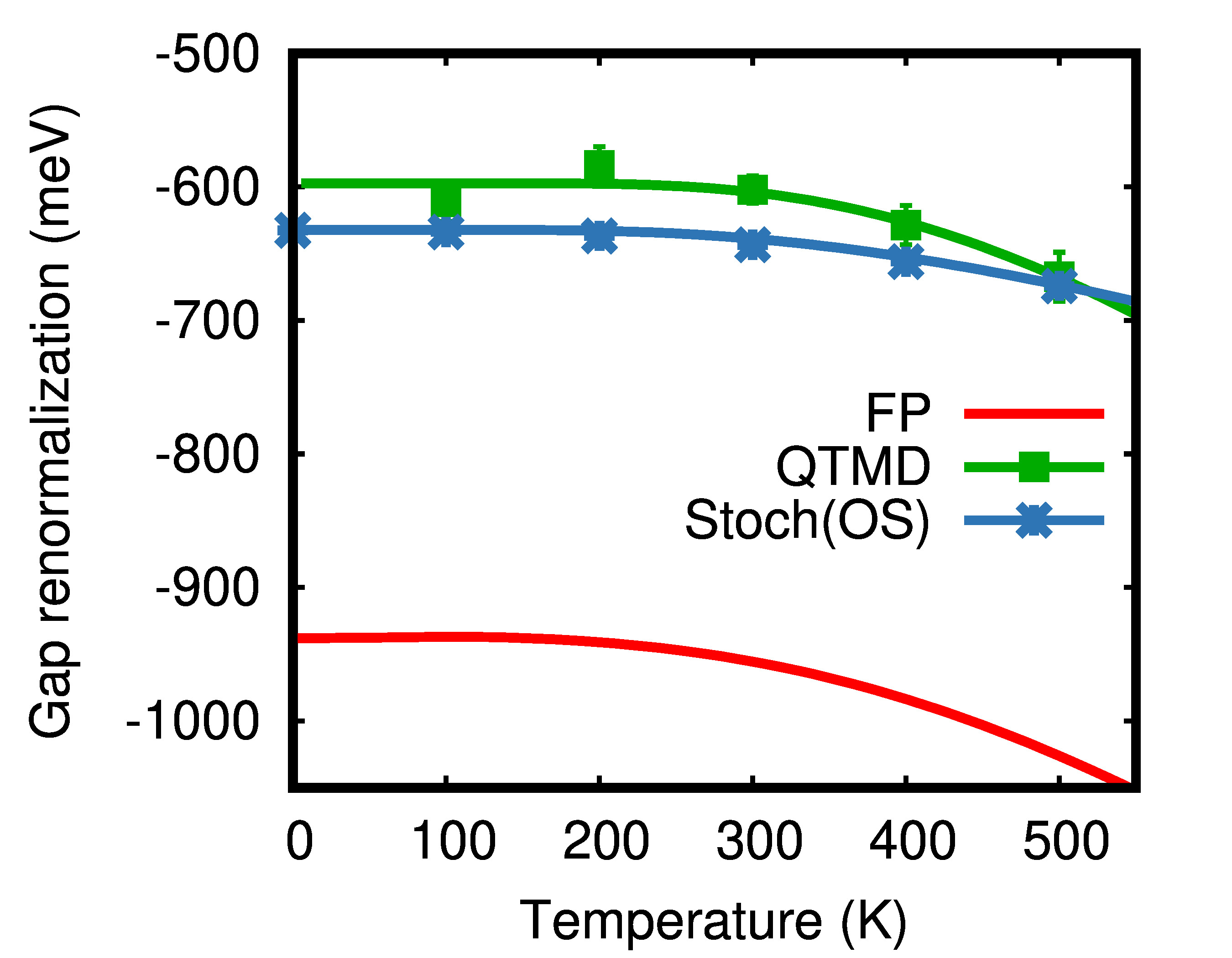}
\caption{Electron-phonon bandgap renormalization of the pentamantane crystal, computed at the SCAN level of theory, as a function of temperature. The gap renormalizations are computed using frozen phonon (FP), stochastic one-shot (Stoch(OS)), and first-principles molecular dynamics coupled with a quantum thermostat (QTMD). For Stoch(OS) and QTMD results, symbols represent the results obtained from simulations while the solid lines represent the Vi\~{n}a model fit of the simulation results.}
\label{fig:penta-cry-gap}
\end{figure}

\begin{figure}[tbhp]
    \centering
    \includegraphics[width=17.7cm]{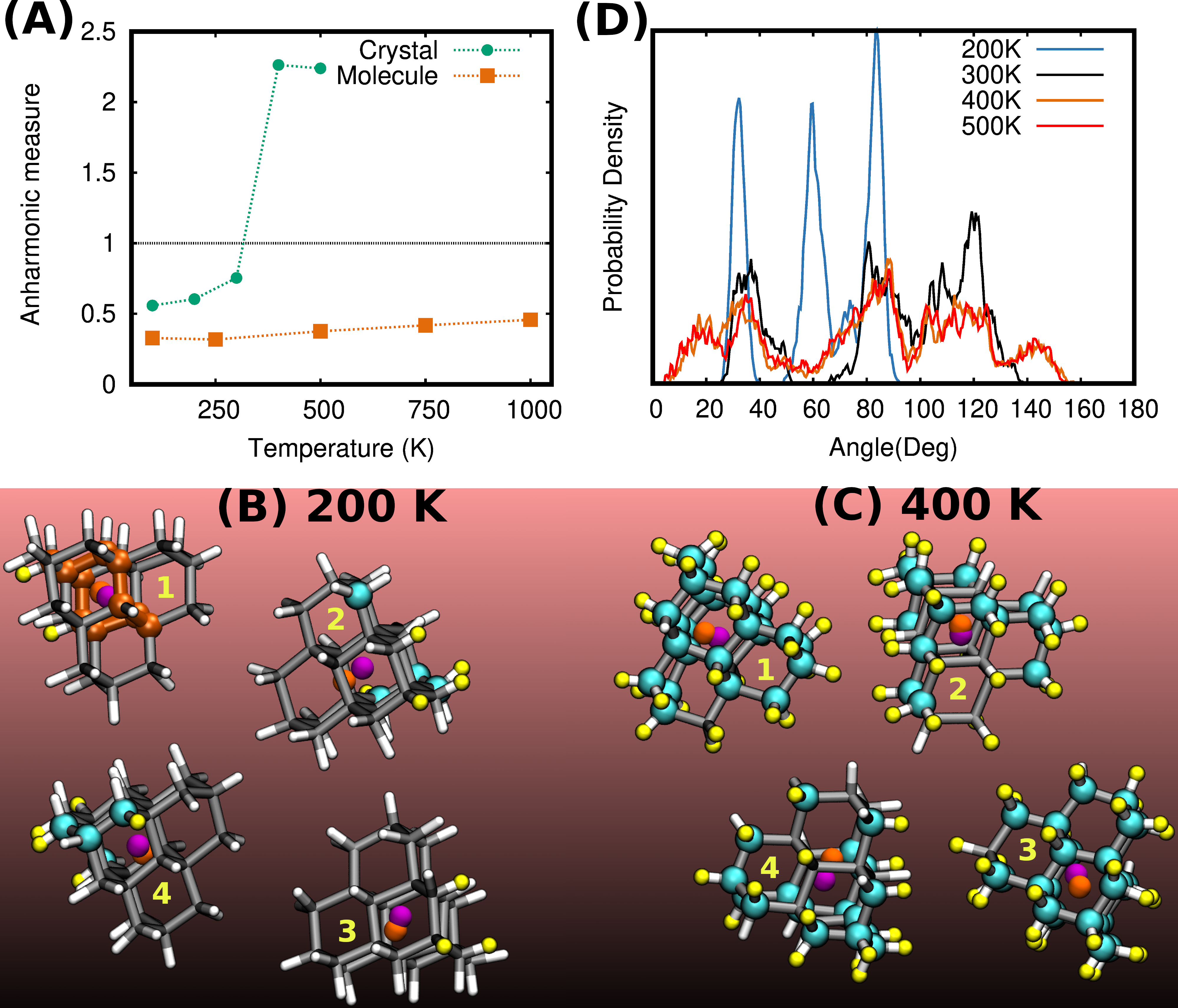}
    \caption{ Panel A: Anharmonic measures of the pentamantane molecule and crystal at different temperatures as obtained from QTMD simulations using the SCAN functional. Panel B-C: The atoms with anharmonic measure greater than 1 are highlighted as cyan (C) and yellow (H) spheres for the pentamantane crystal at 200K and 400K. Within the crystal unit cell, there are four molecules which are designated as 1, 2, 3 and 4. In molecule-1 of panel B, a 10-membered C\textsubscript{10} ring is highlighted with orange balls and sticks. The vector joining the center of mass of the C\textsubscript{10} ring (orange spheres) and that of the molecule (pink spheres) defines a unique molecular axis that is minimally affected by the internal vibration of the molecule. Panel D: The angular distribution between molecular axes at different temperatures as obtained from QTMD simulations.}
    \label{fig:penta-anh-mes}
\end{figure}

Fig. \ref{fig:penta-cry-gap} compares the band gap renormalizations of the pentamantane crystal as a function of temperature when different methods are used to describe the quantum nuclear vibrations. Compared to QTMD results, the FP method overestimates the gap renormalization by more than 300 meV $(>50\%)$, while the stochastic one-shot method yields only a slight $(<6\%)$ overestimate. To understand these differences,  we computed again the crystal's total anharmonic measure using the QTMD trajectories, and our  results are shown in Fig. \ref{fig:penta-anh-mes}A (See Fig S2 in the SI for the mode resolved anharmonic measures at 100 K).     

Though the total anharmonic measure of the crystal is larger than that of the isolated molecule, its value is smaller than one.  At $T > 300$ K, however, the total anharmonic measure becomes larger than 1, and consequently, the harmonic approximation is expected to break down. Yet, the stochastic OS method predicts a ZPR value that is in agreement with that of QTMD. To understand this apparent contradiction, we computed the atom-resolved anharmonic measures. Panels B and C of Fig. \ref{fig:penta-anh-mes} highlight as cyan (C) and yellow (H) spheres those atoms whose anharmonic measure value is greater than one. At 200 K, we found only a handful of such atoms, while at 400 K, most of the atoms have an anharmonic measure value greater than 1. This sharp transition indicates that at $T>300$ K, the crystal's phonon modes become strongly anharmonic due to coupling with rigid body rotations of the molecule. To accumulate further evidence, we defined the axis of a molecular unit by defining a vector joining the centers of mass of a molecule (shown as pink spheres) and the center of mass (shown as orange spheres) of a unique C\textsubscript{10} ring (shown by orange balls and sticks, see molecule-1 of Fig \ref{fig:penta-anh-mes}B) within that molecule. Such a definition of the molecular axis is least sensitive toward intramolecular vibrations. From QTMD trajectories, we computed the probability distribution of the angles between the axes of different molecules within the crystal's unit cell using the TRAVIS trajectory analyzer \cite{travis1,travis2} and these results are shown in panel D of Fig. \ref{fig:penta-anh-mes}. At 200 K, the angular distribution has only three sharp peaks, while at $T>300$ K, a very broad distribution stems from large amplitude rigid body rotations of pentamantane molecules within the crystal. Such hindered rotations have a non-parabolic potential energy surface and are, therefore, strongly anharmonic. However the VBM and CBM states are not localized within the molecules and consequently, they are not much affected by such intermolecular motions. This is the reason why, even at $T>300$ K, the harmonic approximation does not introduce a large error and the stochastic OS results are in good agreement with QTMD simulations.

\begin{figure}[tbhp]
\centering
\includegraphics[width=8.4cm]{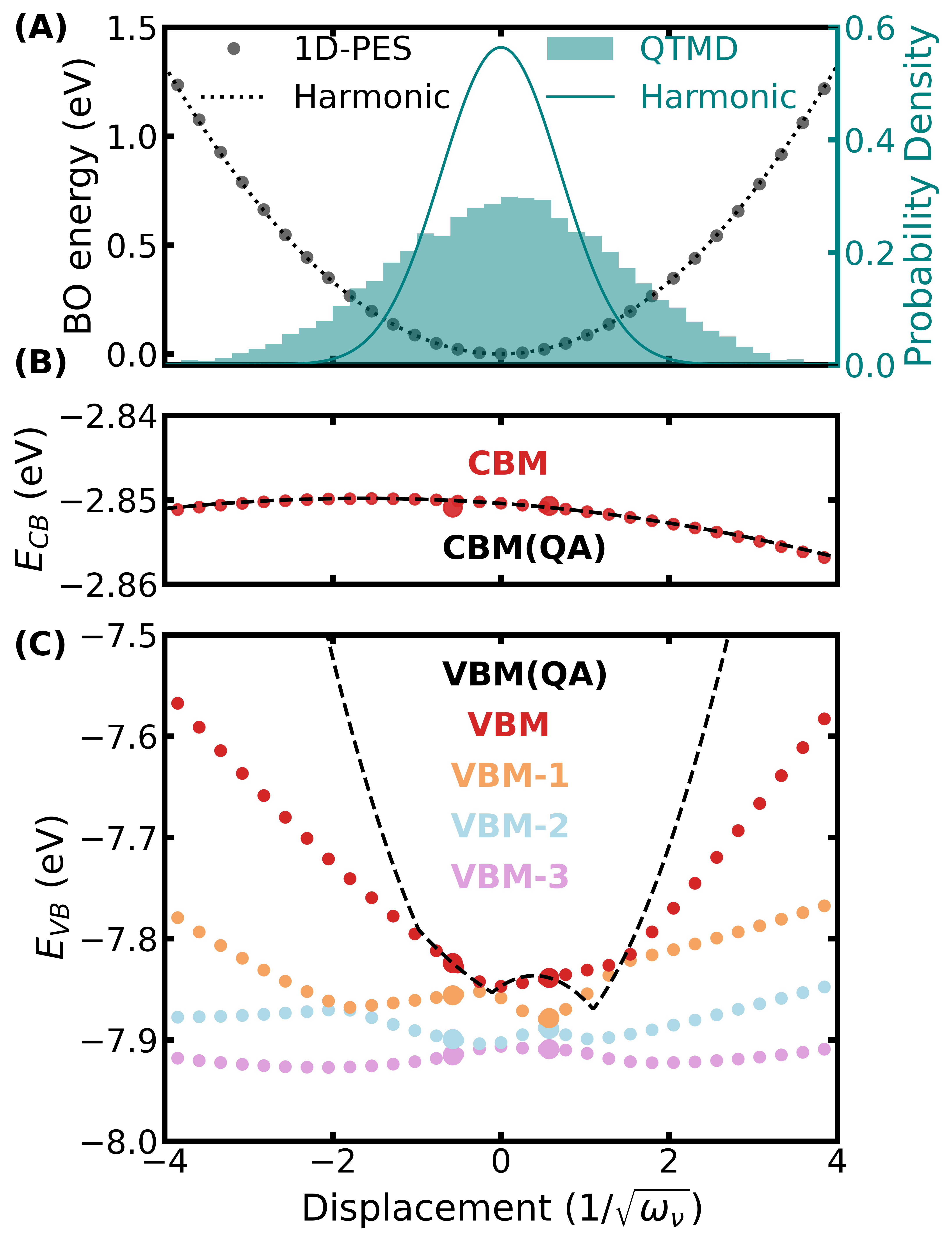}
\caption{One-dimensional scan of the Born-Oppenheimer (BO) potential energy surface (PES), $\V(\bfR)$, and Kohn-Sham eigenenergies ($E_n$) of CBM and VBM along a normal mode with a harmonic frequency of 1336 cm\textsuperscript{-1}. Panel (A) compares the 1D PES computed using SCAN DFT (circles) with that obtained from the harmonic approximation (dotted line). This panel also compares the probability densities at 100 K as obtained from a QTMD simulation and the harmonic approximation. The circles in panels (B) and (C) show the energies of the Kohn-Sham eigenstates of the conduction band  minimum (CBM) and the four highest eigenvalues of the valence band (VB), respectively, as calculated using SCAN DFT. The black dashed line represents the VBM (CBM) eigenenergies computed with the quadratic approximation (QA), using the first and second derivatives of the eigenenergies with respect to the phonon mode. The two points that are used to compute these derivatives are shown with larger symbols.  
}
\label{fig:penta-cry-1d-pes}
\end{figure}

Unlike the stochastic method,  FP calculations overestimate the gap renormalizations by more than 50\% compared to the reference QTMD results, and this large deviation is attributed to the quadratic approximation. To illustrate the importance of non-quadratic electron-phonon coupling terms, we consider the phonon mode along which the electron-phonon coupling energy for the HOMO is the highest when computed using the FP approximation. We  performed a scan of the one-dimensional potential energy surface (PES), and of the Kohn-Sham eigenvalues of the HOMO and LUMO levels along this normal mode. Moreover, we also computed the probability distribution by projecting the XYZ coordinates obtained from QTMD trajectories to the normal mode vector. In Fig \ref{fig:penta-cry-1d-pes}A, we compared the probability density obtained from a QTMD simulation at 100 K with the analytical expression obtained from the harmonic approximation at the same temperature, which is a Gaussian distribution.     

Fig. \ref{fig:penta-cry-1d-pes}A shows that the vibrational energy within the harmonic approximations is in excellent agreement with the Born-Oppenheimer energies obtained by displacing the ions along the normal mode. However, the normal mode probability density obtained with the harmonic approximation is narrower compared to that obtained from the QTMD sampling. This is due to the anharmonic cross-coupling between normal modes which is included in a QTMD sampling but not when the ions are displaced along a normal mode. 
Therefore, sampling a one-dimensional potential energy surface to include anharmonicity, as done in many previous studies \cite{Antonius_PRB_2015, Piccini_Angew_2016, Kundu_JACS_2016} is not necessarily accurate.  
Despite the broadening of the anharmonic vibrational density of the pentamantane crystal, the harmonic approximation to the density is still valid. Therefore, the stochastic method where ions are displaced following the harmonic vibrational density, predicts electron-phonon renormalizations in close agreement with the fully anharmonic QTMD simulations. 

Fig. \ref{fig:penta-cry-1d-pes}B  shows that the CBM eigenenergy surface is flat as a function of phonon mode coordinates, consistent with a small EPCE value (only -0.2 meV) as computed using the FP approach. The CBM eigenenergy surface computed using the quadratic approximation is in excellent agreement with that obtained from DFT with the SCAN functional.
Close to the VBM (see panel C), there are four nearby Kohn-Sham states exhibiting crossing as a function of phonon displacements. Because of these crossings, the VBM state is not always a pure state, and the eigenenergy surface is only a piecewise continuous function with respect to the phonon displacement. Such state crossing introduces additional complexity in using the quadratic approximation, as previously observed also for isolated diamondoid molecules.\cite{FPH_Bester_2019} Therefore, the quadratic approximation overestimates the VBM energies at displacements larger than the value used to compute the first and second derivatives of the eigenenergies. Hence, the FP method predicts a large 72.6 meV contribution toward the ZPR for this phonon mode. In contrast, if we apply stochastic OS displacements only to this mode, the computed ZPR contribution reduces to 20.9 meV. This shows the limitations of the quadratic approximation and of the FP method for the calculations of eigenvalues and energy gaps, and we expect a similar trend for other electronic properties.  

Both the isolated molecule and the crystal of pentamantane have large ZPR values, -500 meV and -600 meV, which are comparable to those of diamond. The frozen-phonon method overestimates the ZPR value of the crystal by 50\% due to the quadratic approximation, whereas the stochastic one-shot method, which does not rely on that approximation, agrees well with the reference QTMD results.

%with EPCEs 72.6, 1.8 meV, 21.7 meV and -66.7 meV, respectively. These states

\subsection{Electron-phonon renormalization of NAI-DMAC molecule and crystal}\label{sec:naidmac}
\begin{figure}[tbhp]
\centering
\includegraphics[width=16.4cm]{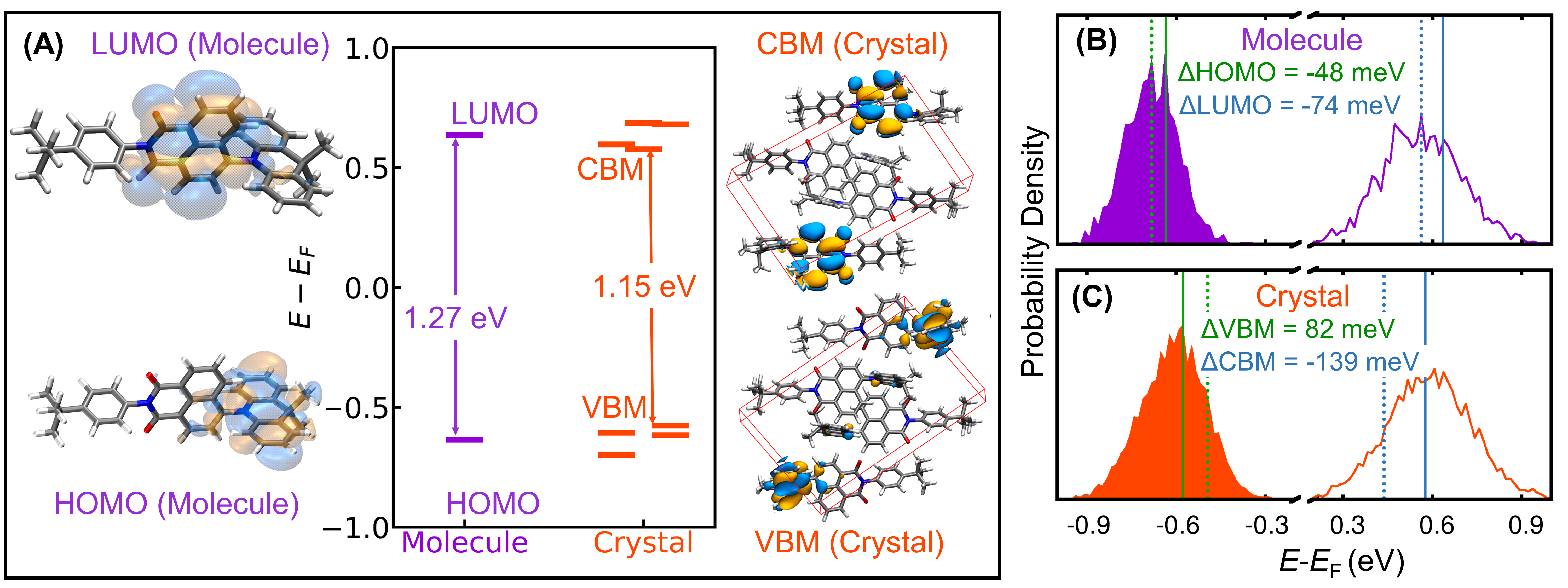}
\caption{Panel A shows the HOMO and LUMO levels relative to the Fermi level computed with the PBE functional for an isolated NAI-DMAC molecule and its molecular crystals when nuclei are at rest. Panels B and C show the electronic density of states at 200 K for the molecule and crystal, respectively, computed using QTMD simulations. The green (blue) vertical lines represent the average energies of the HOMO or VBM (LUMO or CBM) when atoms are at rest (solid line) or in motion at 200 K (broken line). }
\label{fig:naidmac-orbitals}
\end{figure}

We turn to discussing the  results for the NAI-DMAC molecule and crystal. Fig \ref{fig:naidmac-orbitals}A compares the HOMO(LUMO) of the molecule with the VBM(CBM) of the crystal as obtained employing the PBE functional when the nuclei are at rest. The LUMO and HOMO of the molecule, both singly degenerate, are localized on the acceptor unit NAI and donor unit DMAC, respectively. The unit cell of the NAIDMAC crystal has four molecules and 4 Kohn-Sham eigenstates for both CBM and VBM that are very close in energy. The CBM and VBM of the crystal are also localized on the NAI(acceptor) and DMAC (donor) units of two molecules, respectively. With the PBE functional, we obtained a 0.12 eV smaller band gap in the crystal than in the molecule.     

Figs. \ref{fig:naidmac-orbitals}B and \ref{fig:naidmac-orbitals}C show the EDOS for the isolated molecule and the crystal, respectively, when quantum nuclear vibrations at 200 K are included using QTMD simulations. The LUMO(CBM) moves toward the Fermi level by an amount of 74 meV (139 meV); instead,  the HOMO of the molecule moves downwards from the mid-gap by an amount of 48 meV, and the VBM of the crystal towards the Fermi level by an amount of 82 meV.

\begin{figure}[tbhp]
\centering
\includegraphics[width=8.4cm]{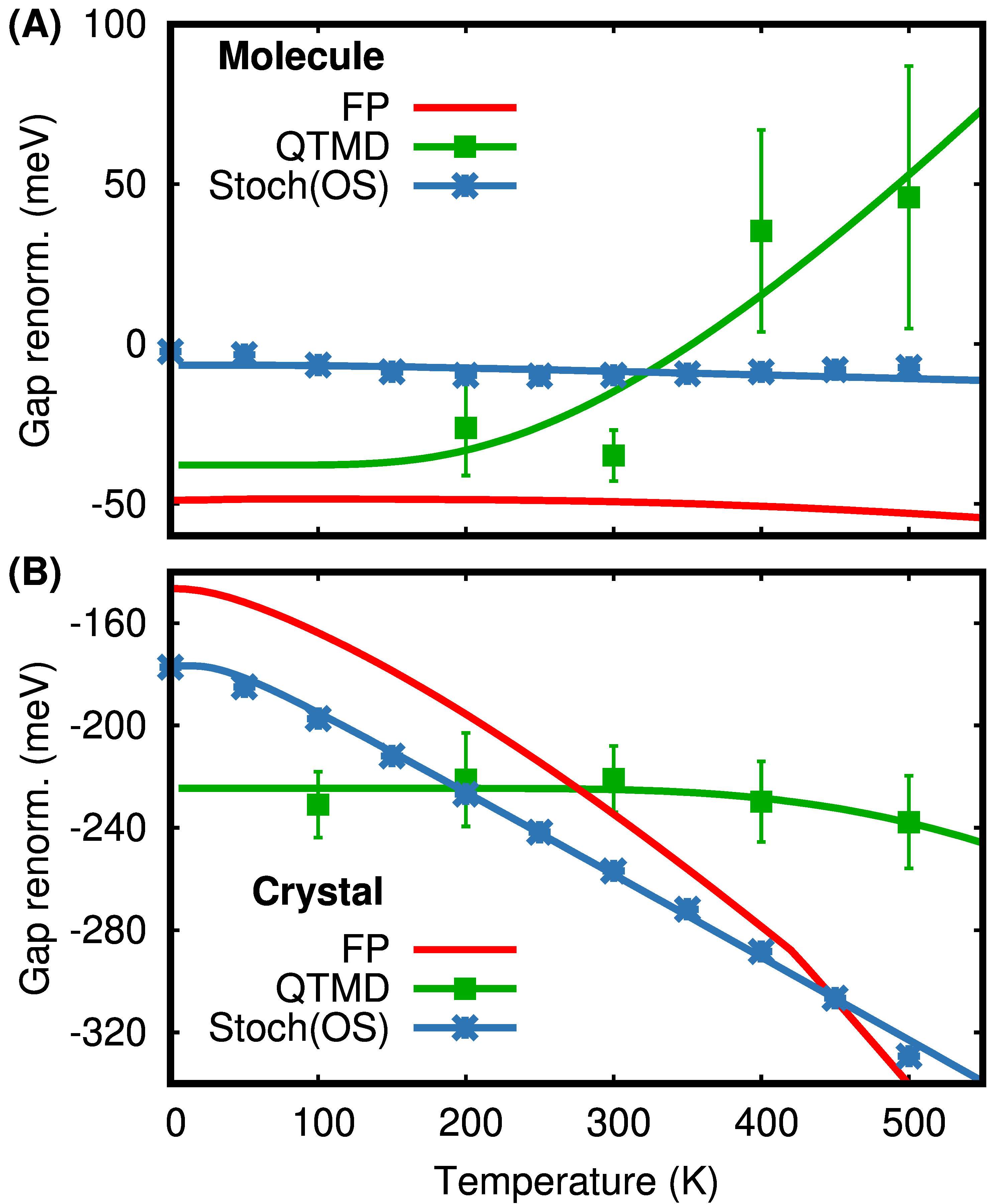}
\caption{Electron-phonon renormalizations of the fundamental gap of the NAIDMAC molecule (panel A) and crystal (panel B) as a function of temperature. The gap renormalizations are computed using the PBE functional and three different approaches: (i) frozen phonon (FP), (ii) stochastic one-shot (Stoch(OS)), and (iii) first-principles molecular dynamics coupled with a quantum thermostat (QTMD). Symbols represent the results of simulations while the solid lines represent the Vi\~{n}a model fit of the simulation results.}
\label{fig:naidmac-gap}
\end{figure}

Fig. \ref{fig:naidmac-gap} compares the fundamental gap renormalizations as a function of temperature for the isolated molecule (panel A) as well as the crystal (panel B) obtained with QTMD simulations, stochastic OS, and FP methods.  For the isolated molecule, surprisingly, the gap renormalization values are rather small, $\simeq -40$ meV below 300 K, however, they change sign (i.e. the gap increases) when the temperature is above 400 K and the slope of the gap renormalization with respect to temperature is positive. Indeed, with increasing T, the HOMO level decreases in energy while the LUMO level remains approximately constant. In contrast, for the crystal, up to almost 500 K, the band gap renormalizations remain constant at $\simeq$ -220 meV with very small variations observed for the CBM and VBM. 

\begin{figure}[tbhp]
\centering
\includegraphics[width=16cm]{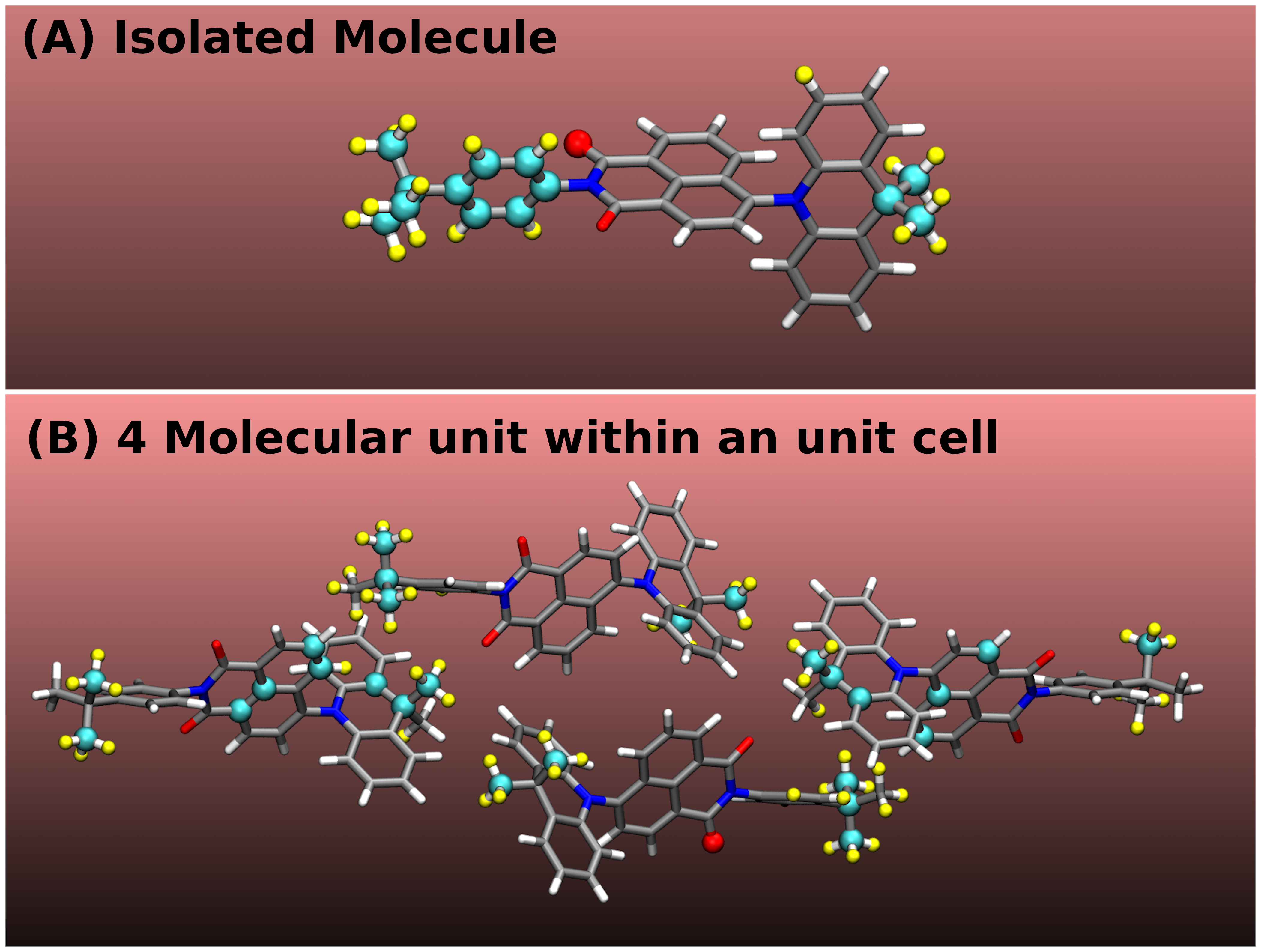}
\caption{Atom resolved anharmonic measure of an isolated NAI-DMAC molecule (panel A) and the crystal (panel B). The atoms with an anharmonic measure greater than 1 are highlighted as cyan (C) and yellow (H) spheres.}
\label{fig:naidmac-anharm-mes}
\end{figure}

For both the molecule and the crystal, the FP and stochastic methods predict an incorrect temperature dependence of the fundamental gap, and at variance with the pentamantane crystal's results, we find that the stochastic method does not improve the FP results. We find that the anharmonic measure from the QTMD simulations is 7.29 units and 2.99 units for the molecule and the crystal, respectively, even at a very low temperature of 200 K. (Figures S3-S4 in the SI show the mode resolved anharmonic measures of the molecule and the crystal). There are several modes with very high values of the anharmonic measure (in the range of 10-50) showing that the harmonic approximation breaks down. Fig. \ref{fig:naidmac-anharm-mes} shows the atoms for which the anharmonic measure value is larger than 1 for the molecule (panel A) and the crystal (panel B). We note that for the isolated molecule, the atoms of the tertiary-butyl-phenyl group of the NAI subunit and the methyl groups of the DMAC subunit exhibit a large anharmonic measure,  suggesting a likely free rotation of these groups. 

\begin{figure}[tbhp]
\centering
\includegraphics[width=8.4cm]{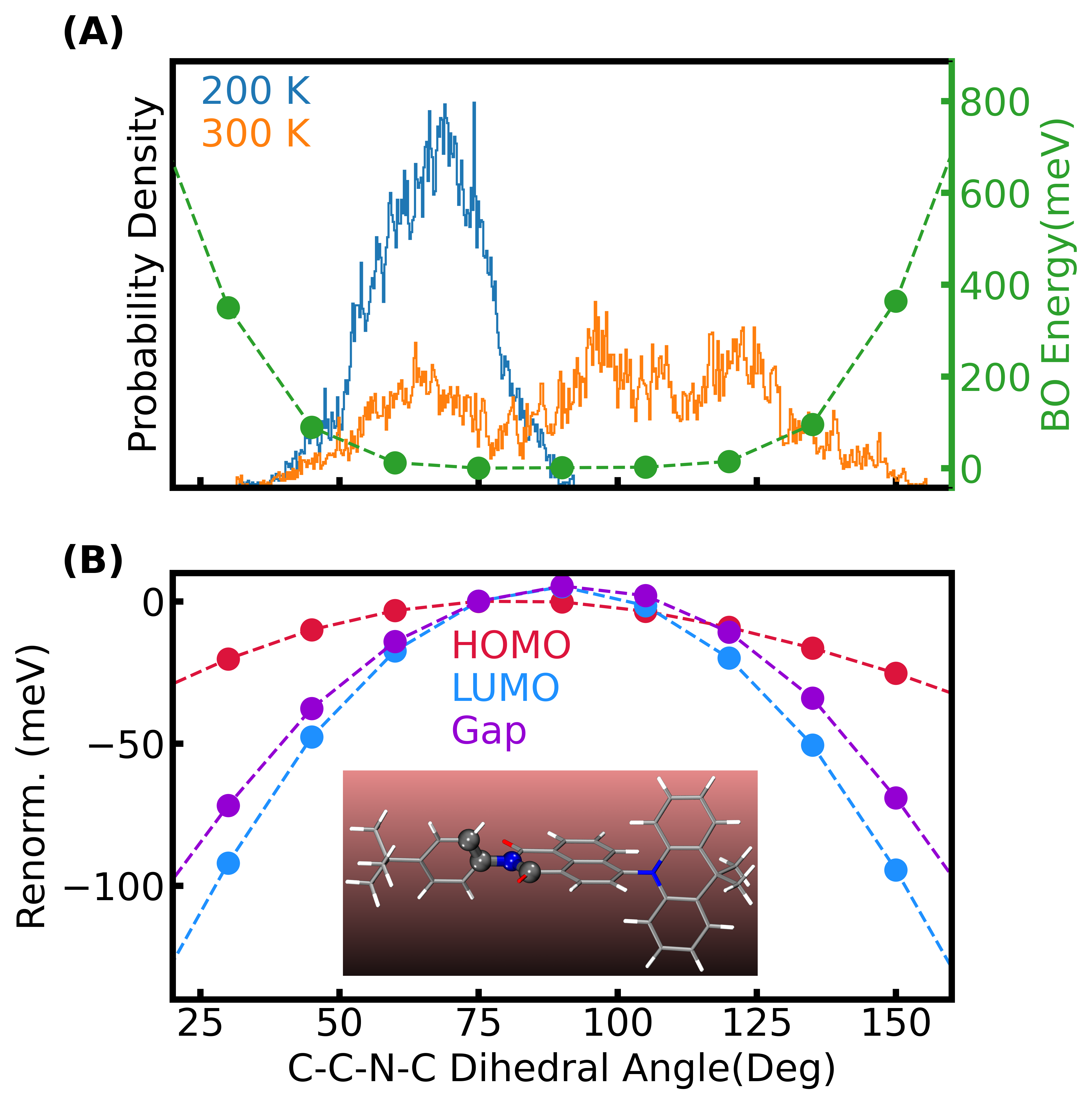}
\caption{Panel A: Probability distribution of the C-C-N-C dihedral angle (highlighted as spheres in the inset of panel B) for the isolated molecule at different temperatures as obtained from QTMD simulations. The green circles represent the Born-Oppenheimer energy when PES is scanned by varying the dihedral angle of the isolated molecule. Panel B: The renormalizations for the HOMO, and LUMO levels as well as the gap for the isolated molecule when only the dihedral angle is varied. }
\label{fig:naidmac-mol-dihed-scan}
\end{figure}

\begin{figure}[tbhp]
\centering
\includegraphics[width=8.4cm]{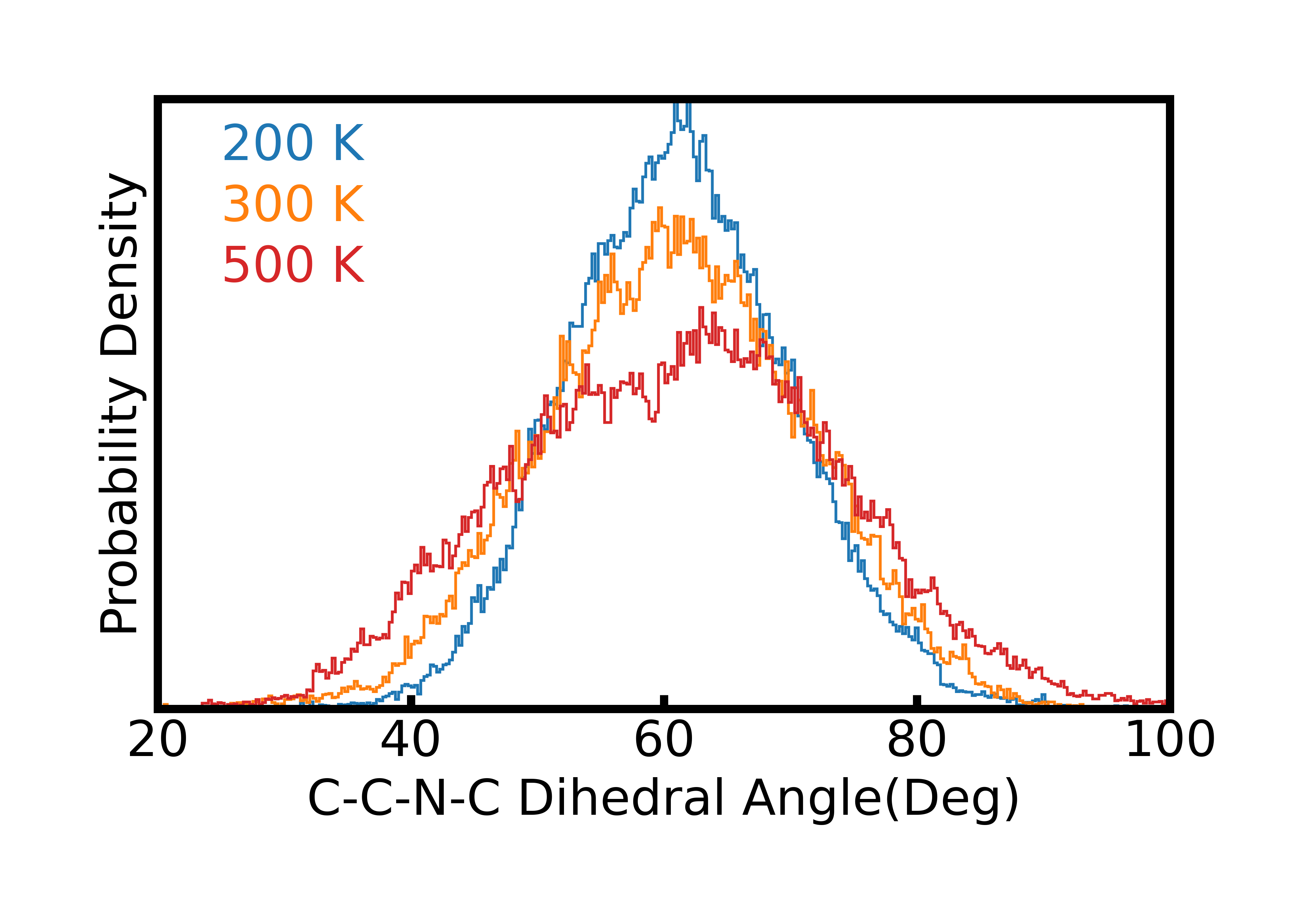}
\caption{Probability distribution of the C-C-N-C dihedral angle (highlighted as spheres in the inset of Fig. \ref{fig:naidmac-mol-dihed-scan}B) for the NAI-DMAC crystal at different temperatures, as obtained from QTMD simulations.}
\label{fig:naidmac-cry-dihed-prob}
\end{figure}

 To understand the effects of free rotations, we scanned the PES and HOMO, LUMO eigenenergies by rotating the tertiary-butyl-phenyl group relative to the naphthalimide group. We note that such a one-dimensional scan neglects the coupling between these rotations with other rotational and/or vibrational degrees of freedom and such a one-dimensional PES scan should only be used to gain a qualitative understanding. Fig. \ref{fig:naidmac-mol-dihed-scan}, shows the BO energies (panel A) and renormalization of the HOMO, LUMO energies  and their gap as a function of C---C---N---C dihedral angle.
In addition, the probability distributions of this dihedral angle at different temperatures as obtained from QTMD simulations are reported for the molecule (Fig. \ref{fig:naidmac-mol-dihed-scan}A) and the crystal (Fig. \ref{fig:naidmac-cry-dihed-prob}).  

In the isolated molecule, the rotation of the tertiary-butyl-phenyl group is characterized by a very flat PES, see Fig. \ref{fig:naidmac-mol-dihed-scan}A. Although at 200 K, the dihedral angle is restricted within 40\degree---100\degree, at $T>300$ K, the rotation becomes almost free, and a wide range of dihedral angles ( 25\degree --- 160\degree) are explored by the torsional motion. 
Both LUMO and HOMO levels are coupled to this rotational degree of freedom and their energy decreases as the dihedral angle moves further away from its equilibrium value ($\simeq 75$\degree); however, the energy decrease is faster for the LUMO than the HOMO and the resulting gap renormalization is negative. 
 We note that the LUMO and HOMO are strongly coupled to the rotation of the NAI unit relative to the DMAC unit as well (see Figure S5 in SI), but the former increases in energy while the latter decreases in energy as the C---C---N---C dihedral angle moves away from its equilibrium value ($\simeq 90$\degree), and consequently, the gap opens with this rotation.    
We also scanned the PES for the rotation of the methyl groups, but the effects on the HOMO-LUMO gap are much smaller, 5 meV at most, for a methyl group; see Figs. S6-S8 in the SI. 
Such torsional and rotational motions are strongly anharmonic because (i) the PES is sinusoidal and hence cannot be modeled with a parabola, and (ii) curvilinear rotational displacements cannot be described by rectilinear normal modes when the angle variation is large. 

In the crystal, contrary to the isolated molecule, the torsional motion of the tertiary-butyl-phenyl group is restricted due to packing constraints. The dihedral angle distribution varies weakly  with increasing temperature and, the carbon atoms of the phenyl rings of the tertiary-butyl-phenyl group do not exhibit a large anharmonic measure. Despite the packing constraints, the methyl groups are still free to rotate and consequently, their motions remain strongly anharmonic. Therefore, the strong anharmonic phonon-phonon interactions originating within an isolated NAI-DMAC molecule also persist within the molecular crystal, and these interactions cannot be accurately represented using the stochastic method.

\begin{figure}[tbhp]
\centering
\includegraphics[width=8.4cm]{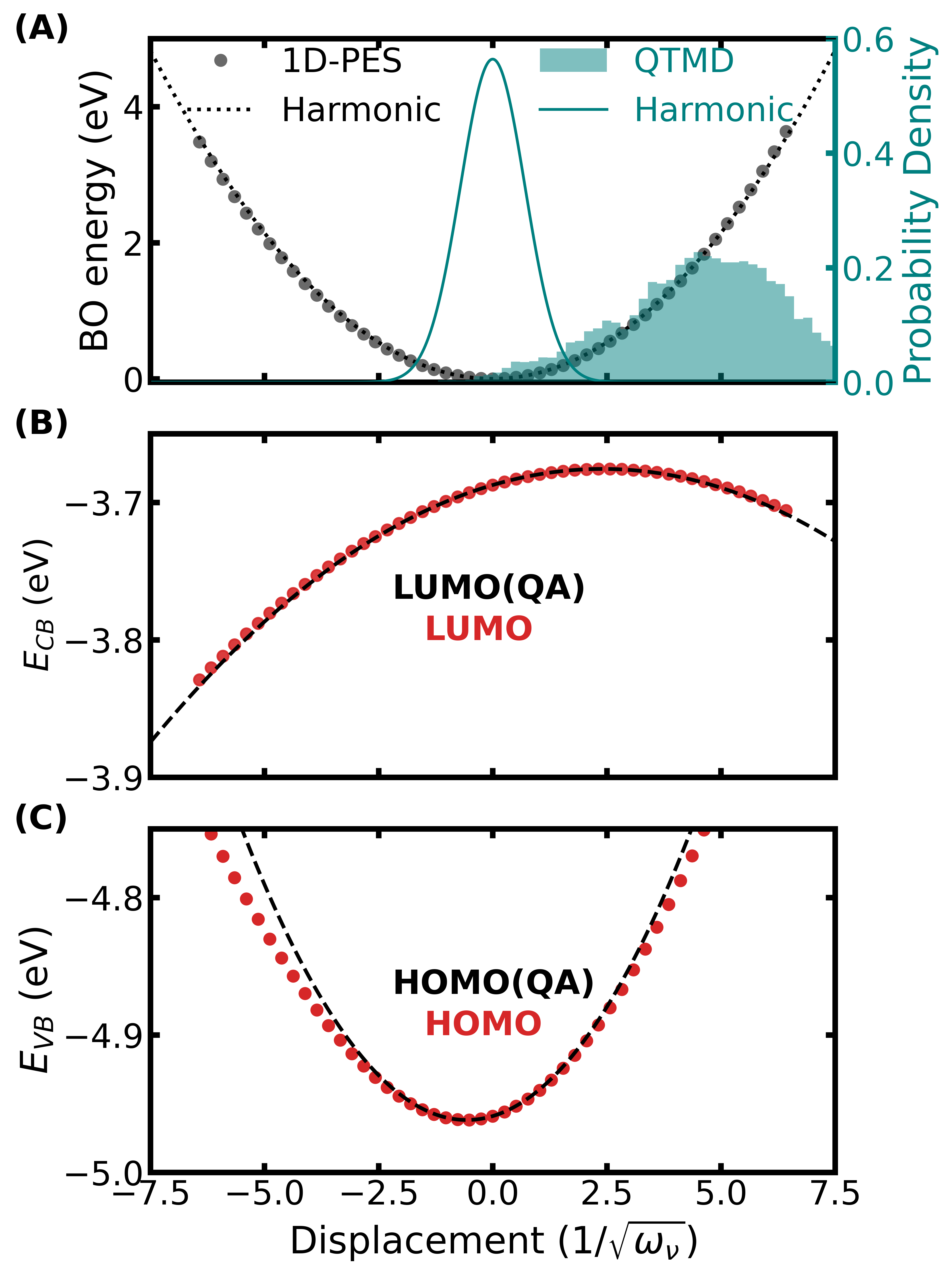}
\caption{One-dimensional scan of the Born-Oppenheimer (BO) potential energy surface (PES), $\V(\bfR)$, and Kohn-Sham eigenenergies ($E_n$) of CBM and VBM along a normal mode with a harmonic frequency of 1329 cm\textsuperscript{-1}. Panel (A) compares the 1D PES computed using PBE (circles) with that obtained from the harmonic approximation (dotted line). This panel also compares the probability densities at 200 K as obtained from a QTMD simulation and the harmonic approximation. The circles in panels (B) and (C) show the energies of the Kohn-Sham eigenstates of the lowest unoccupied molecular orbital (LUMO) and the highest occupied molecular orbital (LUMO), respectively, as calculated using DFT with PBE functional. The black dashed line represents the HOMO (LUMO) eigenenergies computed with the quadratic approximation (QA), using the first and second derivatives of the eigenenergies with respect to the normal mode.}
\label{fig:naidmac-mol-1d-pes}
\end{figure}

\begin{figure}[tbhp]
\centering
\includegraphics[width=8.4cm]{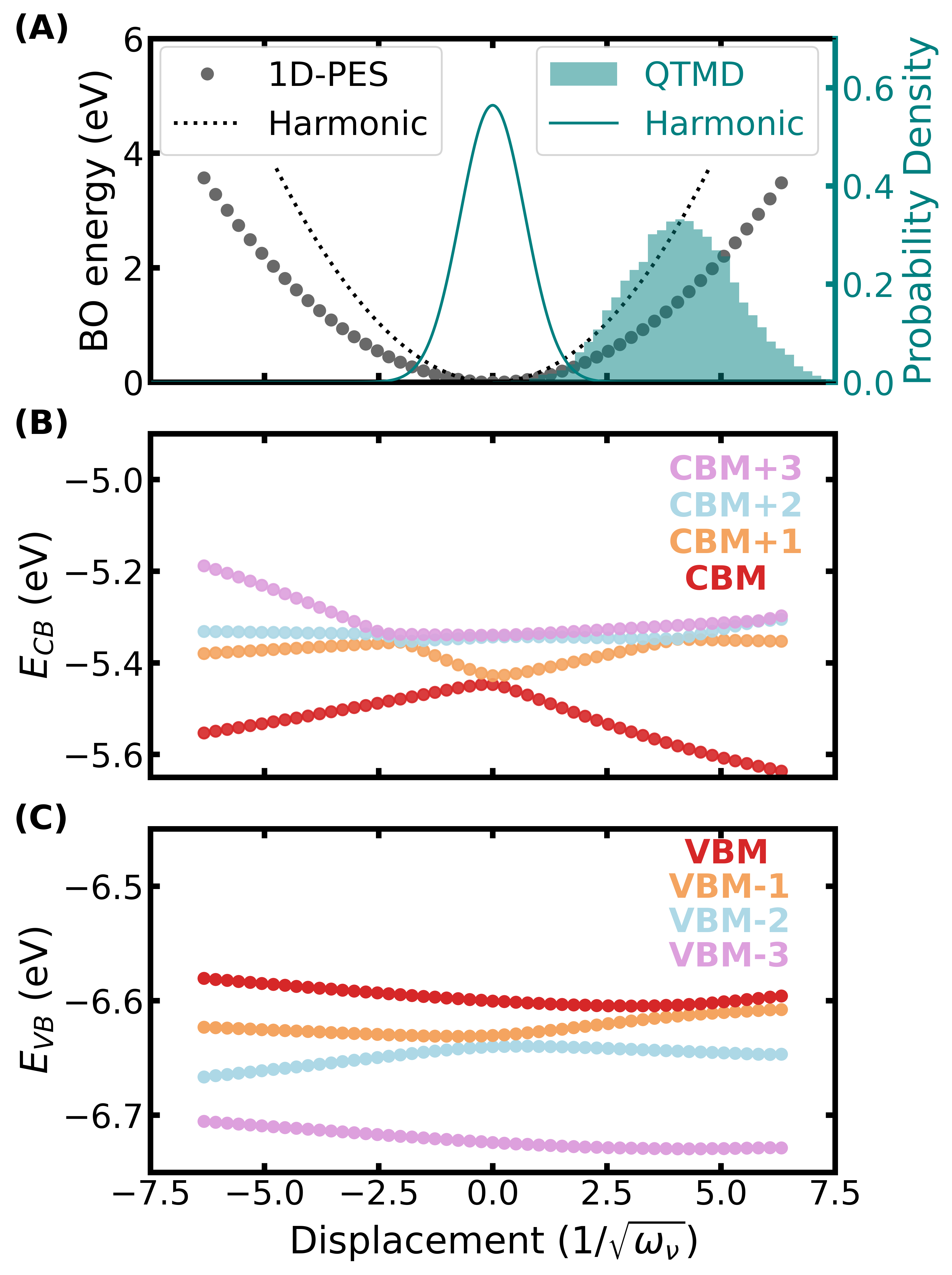}
\caption{
One-dimensional scan of the Born-Oppenheimer (BO) potential energy surface (PES), $\V(\bfR)$, and Kohn-Sham eigenenergies ($E_n$) of CBM and VBM along a normal mode with a harmonic frequency of 1371 cm\textsuperscript{-1}. Panel (A) compares the 1D PES computed using PBE (circles) with that obtained from the harmonic approximation (dotted line). This panel also compares the probability densities at 200 K as obtained from a QTMD simulation and the harmonic approximation. The circles in panels (B) and (C) show the energies of the Kohn-Sham eigenstates of the conduction band (CB) with the four lowest eigenvalues (CBM, to CBM+3) and the valence band (VB) with the four highest eigenvalues (VBM to VBM-3) respectively, as calculated using DFT with the PBE functional. 
%The black dashed line represents the HOMO (LUMO) eigenenergies computed with the quadratic approximation (QA), using the first and second derivatives of the eigenenergies with respect to the phonon mode.
}
\label{fig:naidmac-cry-1d-pes}
\end{figure}

To understand the consequences of the (i) harmonic and (ii) quadratic approximations, we considered the phonon modes with the highest electron-phonon coupling energy for the molecule's HOMO and the crystal's CBM, respectively. We  performed a scan of the one-dimensional potential energy surface (PES), and Kohn-Sham eigenvalues of the HOMO(VBM) and LUMO(CBM) levels along this normal mode (see Figs. \ref{fig:naidmac-mol-1d-pes} and \ref{fig:naidmac-cry-1d-pes}). In addition, we also computed the vibrational density of these normal modes by projecting the XYZ coordinates obtained from QTMD trajectories to the normal mode vectors. In Figs. \ref{fig:naidmac-mol-1d-pes}A and \ref{fig:naidmac-cry-1d-pes}A, we compared such a vibrational density obtained from a QTMD simulation at 200 K with the analytical expression obtained from the harmonic approximation at the same temperature, which is a Gaussian distribution.

First, we consider the isolated molecule. Fig. \ref{fig:naidmac-mol-1d-pes}A shows again an excellent agreement between the harmonic potential energy and the Born-Oppenheimer energies obtained by displacing the ions along the normal mode. We note that the estimation of the anharmonic measure for this mode as obtained from QTMD simulation at 200 K (3.02) is larger than one, pointing at the inadequacy of the normal mode to describe the inter-nuclear motion within the molecule. This stems from a strong anharmonic coupling between the chosen normal mode with other ones, resulting in an asymmetric vibrational density along this mode. This observation reiterates the limitations of the one-dimensional scans along normal modes to include anharmonic coupling effects. 
Fig. \ref{fig:naidmac-mol-1d-pes}B and C show the LUMO and HOMO energies as a function of normal mode displacements, respectively. In both cases, we observe a reasonable agreement between the results obtained using the quadratic approximation and those of the first-principles calculations. 
%Both the stochastic and FP methods rely on the Gaussian approximation for the vibrational density that originates from the harmonic approximation, while the latter additionally employs quadratic approximation to simplify the thermal averaging procedure. 
As the anharmonic vibrational density for the NAI-DMAC molecule is not a Gaussian distribution, both FP and stochastic methods become inaccurate to compute the electron-phonon renormalization of the band gap. 

Fig. \ref{fig:naidmac-cry-1d-pes}A shows that the harmonic approximation overestimates the vibrational energy compared to the reference first principle calculations when ions are displaced along the normal mode. We note that the true anharmonic vibrational density is not a Gaussian distribution and it is asymmetric and shifted towards a  positive value of $x$.
 Fig. \ref{fig:naidmac-cry-1d-pes}B and C show the energies of the Kohn-Sham eigenstates with four lowest and highest eigenvalues of the conduction and valence bands respectively, as a function of normal mode displacements.  As for the pentamantane crystal, we observe crossings between valence and conduction band states which make the quadratic approximation unreliable for the electronic states. 
Therefore, both FP and stochastic methods predict inaccurate results for the electron-phonon renormalization of the bandgap as neither the quadratic nor the harmonic approximation hold for the NAI-DMAC molecular crystal. 

%For both the isolated molecule and the crystal of NAI-DMAC, neither the frozen-phonon nor the stochastic one-shot method corroborates well with the reference QTMD results. This is because of the presence of several torsional modes within the molecule makes the molecule very anharmonic. Consequently, vibrational density is not anymore a Gaussian distribution as predicted by the harmonic approximation. Both the frozen-phonon and stochastic method relies on Gaussian approximation for the vibrational density and therefore, both methods fail to compute electron-phonon renormalization accurately. Though the packing constraint within the crystal makes the torsional motion more restricted than an isolated molecule, this is not sufficient for the success of harmonic approximation. Therefore the intra-molecular anharmonicity originating from the molecule persists in the crystal and neither the stochastic nor the frozen-phonon approach is trustworthy.      

\section{Summary and Conclusions}\label{sec:conclusion}

We investigated the effect of quantum vibronic coupling and anharmonicity on the electronic properties of two molecular crystals, pentamantane, composed of rigid molecules, and NAI-DMAC, composed of floppier units. Specifically, we analyzed the validity of common approximations used in the literature when applying the stochastic and frozen phonon methods, by comparing the results of these techniques with those of first principles molecular dynamics simulations with a quantum thermostat.
We found that for both the isolated diamondoid molecule and corresponding molecular crystal the ZPR of their electronic gaps is rather large, -500 meV and -600 meV (at the SCAN level) respectively, comparable to that of diamond. In contrast, the NAI-DMAC molecule has a small ZPR of -38 meV (at the PBE level), despite consisting of light first and second-row elements. Due to packing constraints, the molecular crystal of NAI-DMAC has a relatively larger ZPR of -224 meV.

Our calculations showed that in spite of PBE and SCAN functionals giving a similar anharmonic measure for an isolated diamondoid molecule,  results with the SCAN functional exhibit  more pronounced higher-order electron-phonon coupling effects. In fact, the quadratic approximation and resulting frozen phonon results are poorer approximations when using SCAN, which includes van der Waals interactions.

 We also found that the presence of nearly degenerate states close to the valence band maxima (conduction band minima) of the molecular crystals studied here made the frozen phonon calculations inaccurate, leading to incorrect predictions  of the VBM, CBM and gap renormalizations. 
A stochastic calculation may improve over the frozen phonon methods, provided the anharmonic phonon-phonon scattering is relatively weak.
However, when the anharmonic phonon-phonon scattering is substantial,  then  both frozen phonon and stochastic methods fail to appropriately describe electron-phonon renormalizations. A strong anharmonic coupling is expected for floppy molecules and their molecular crystals and the frozen phonon or stochastic method should not be used for such systems.

Finally, we found that sampling one-dimensional phonon modes to incorporate anharmonic effects, as in many earlier studies, may be inaccurate, as it fails to account for  anharmonic couplings between normal modes. Our work highlights the importance of including quantum vibronic effects to accurately describe the electronic properties of molecular crystals composed of light elements.

%\section{Associated Contents}
%\subsection{Supporting Information}

%The Supporting information is available free of charge at

%S1. Methods. S1.1. Theory, S1.2 Stochastic Approach, S1.3 Frozen Phonon Approach.
%S2. Additional Figures. S2.1 Mode Resolved EPCEs and Anharmonic Measures, S2.2 Effect of torsion on the potential energy surface and band gap of NAI-DMAC molecule. 

%\section{Authors}
%{\bf Corresponding Authors}

%rpan Kundu \\ Pritzker School of Molecular Engineering, The University of Chicago, Chicago, Illinois, 60637, United States; \\ https://orcid.org/0000-0001-5351-3254; \\ Email: arpan.kundu@gmail.com

%Giulia Galli \\ Pritzker School of Molecular Engineering, The University of Chicago, Chicago, Illinois, 60637, United States; \\ Department of Chemistry, University of Chicago, Chicago, Illinois 60637, United States; \\ Materials Science Division and Center for Molecular Engineering, Argonne National Laboratory, Lemont, Illinois 60439, United States; \\ https://orcid.org/0000-0002-8001-5290; \\ Email: gagalli@uchicago.edu

\begin{acknowledgement}
We thank Dr. Tommaso Francese for many useful discussions.
This work was supported by MICCoM, as part of the Computational Materials Sciences Program funded by the U.S. Department of Energy, Office of Science, Basic Energy Sciences, Materials Sciences, and Engineering
Division through Argonne National Laboratory, under Contract No. DE-AC02-06CH11357. This research used resources of the University of Chicago Research Computing Center.
 
\end{acknowledgement}

%\bibliography{references}
\providecommand{\latin}[1]{#1}
\makeatletter
\providecommand{\doi}
  {\begingroup\let\do\@makeother\dospecials
  \catcode`\{=1 \catcode`\}=2 \doi@aux}
\providecommand{\doi@aux}[1]{\endgroup\texttt{#1}}
\makeatother
\providecommand*\mcitethebibliography{\thebibliography}
\csname @ifundefined\endcsname{endmcitethebibliography}
  {\let\endmcitethebibliography\endthebibliography}{}

%\clearpage
%\section{TOC figure}
%\begin{figure}
%    \centering
%    \includegraphics[width=\linewidth]{toc.png}
%    \caption{Table of content figure}
%\end{figure}

\end{document}

\maketitle

\tableofcontents

\section{Methods}\label{sec:method}
\subsection{Theory}
Within the Born-Oppenheimer (BO) approximation, the Schr\"odinger equation for the quantum nuclear  motion of a system with $N$ nuclei can be expressed as\cite{Monserrat_Rev_2018},
\begin{equation} \label{eq:vib_se}
\h |\chi_k(\bfR) \rangle = \qty(-\frac{1}{2}\sum_{I}\frac{1}{M_I}\nabla_{R_I}^2 + \V(\bfR)) |\chi_k(\bfR) \rangle = \varepsilon_k |\chi_k(\bfR) \rangle,
\end{equation}
where $M_I$ and $\bfR = (R_{1x},R_{1y},...,R_{Nz})$ are the mass of the $I$-th nucleus and the Cartesian position vector of all nuclei with respect to a chosen origin, respectively. $\V(\bfR)$, and $|\chi_k(\bfR)\rangle$ represent the $3N$-dimensional adiabatic potential energy surface, and the wave function for the k-th nuclear state, respectively. Since, $\V(\bfR)$ and $|\chi_k(\bfR)\rangle$ are indifferent to the choice of the origin, for the convenience of discussion, we set the origin at the coordinates of the equilibrium geometry where $\V(\bfR)$ is minimum. We consider only the $\Gamma$-point vibrations within the simulation cell. 

We consider an electronic observable, e.g., electronic eigenenergies or total energy, $\aobs(\bfR) = \langle \psi(\bfr; \bfR) | \hat{\aobs}(\bfr; \bfR) | \psi(\bfr; \bfR) \rangle $, that depends on one electronic state. 
Here $\bfr$  and $|\psi(\bfr; \bfR) \rangle$ are the electronic coordinates and wave function, respectively. We note that as a consequence of BO approximation, the electronic wave function has only parametric dependence on the nuclear coordinates, $\bfR$. 
When the system is at equilibrium at temperature $T$,  the effect of electron-phonon interaction on the electronic property $\aobs(\bfR)$ can be included by performing an ensemble average over all adiabatic nuclear states $k$,
\begin{equation} \label{eq:ensemble_avg}
\begin{aligned}
\langle \aobs \rangle_T = \frac{1}{Q(T)}\sum_{k=0}^{\infty} \langle\chi_k(\bfR) | \hat{\aobs}(\bfR) |\chi_k(\bfR) \rangle \exp(-\frac{\varepsilon_k}{k_BT}) = \int d\bfR W(\bfR,T) \aobs (R).
\end{aligned}
\end{equation}
Here,
\begin{equation}
    W(\bfR,T) = \frac{1}{Q(T)} \sum_{k=0}^{\infty}\langle \chi_k(\bfR) | \chi_k(\bfR) \rangle \exp(-\frac{\varepsilon_k}{k_BT}),
\end{equation}
denotes the probability of finding the system with the nuclear coordinates within $\bfR$ and $\bfR+d\bfR$.
The partition function $Q$ is defined as,
\begin{equation} \label{eq:partion_fn}
Q(T) = \sum_{k=0}^{\infty} \exp(-\frac{\varepsilon_k}{k_BT}),
\end{equation}
where $k_B$ is the Boltzmann constant. 

A molecular dynamics simulation with either a path-integral approach\cite{PI_Berne_Rev_1986,PI_hererro_rev_2014} or quantum thermostat approach \cite{QT_Ceriotti_PRL_2009,QT_Ceriotti_JCTC_2010,QT_Review_Finocchi_2022} utilizes Eq. \ref{eq:ensemble_avg} to compute the electron-phonon renormalized electronic properties. However,  being computationally expensive, such simulations are rarely adopted for computing electron-phonon renormalizations from first principles. The standard approaches in solid-state physics employ approximations to $\V(\bfR)$, to simplify the expression in Eq. \ref{eq:ensemble_avg}.

A Taylor series expansion of $\V(\bfR)$ near the equilibrium geometry, $\bfR = \bfzero$, and subsequently, neglecting the terms beyond second order yields the potential energy with harmonic approximation (HA),
\begin{equation} \label{eq:ha}
\V^{\ha}(\bfR) = \frac{1}{2}\sum_{I\alpha,J\alpha'}(R_{I\alpha}\sqrt{M_I})D_{I\alpha,J\alpha'}(\sqrt{M_J}R_{J\alpha'}) = \frac{1}{2}\bfR^{T}\bfM^{1/2}\bfD\bfM^{1/2}\bfR,
\end{equation}
with $\bfM$ representing a $3N\times3N$ diagonal matrix of nuclear masses. We note that the first order term of the Taylor expansion becomes zero by setting the condition of the minimum, $\pdv{\V(\bfR)}{R_{I\alpha}}=0$. In addition, we set the reference potential energy, $\V(\bfzero)=0$. The elements of the dynamical matrix, which is also known as the mass-weighted Hessian matrix is given by,
\begin{equation} \label{eq:dynmat}
      D_{I\alpha,J\alpha'} = \frac{1}{\sqrt{M_{I}M_{J}}}\eval{\pdv{\V(\bfR)}{R_{I\alpha}}{ R_{J\alpha'}}}_{\bfR=\bfzero},
\end{equation}
where $\alpha$,$\alpha'$ denote the cartesian axes $x$,$y$ or $z$ and $I,J$ denote the indices of the nuclei. Spectral decomposition of the dynamical matrix, $\bfD = \bfU\bfOmega^2\bfU^T$, returns a unitary matrix, $\bfU$, and a $3N\times3N$ diagonal matrix of normal-mode frequencies, $\bfOmega$, with diagonal elements: $\omega_1,\omega_2,...,\omega_{3N}$. The unitary matrix, $\bfU$, defines the cartesian to normal mode transformations,
\begin{equation}\label{eq:cart2norm}
\begin{aligned}
    \bfX=\bfU^T\bfM^{1/2}\bfR, \\
     \nabla_X^2=\bfU^T\bfM^{1/2}\nabla_R^2,
\end{aligned}
\end{equation}
and back transformation to cartesian from normal modes,
\begin{equation}\label{eq:norm2cart}
\begin{aligned}
    \bfR=(\bfM^{1/2})^{-1}\bfU\bfX, \\
     \nabla_R^2=(\bfM^{1/2})^{-1}\bfU\nabla_X^2,
\end{aligned}
\end{equation}
with $\bfX$ representing the matrix of $3N$ normal mode vectors.

After transformation to normal modes, the total nuclear Hamiltonian separates into 3 independent translational degrees of freedom, $d_r$ number of independent global rotational degrees of freedom, and $3N-3-d_r$ number of independent vibrational degrees of freedom. For a solid, linear isolated molecule, and a non-linear isolated molecule, the number of rotational degrees of freedom ($d_r$) is 0, 2, and 3, respectively. Since the translations and global rotations, which usually appear as the first $3+d_r$ lowest eigenvalues,  do not affect the electronic properties, we focus on the vibrational Hamiltonian which becomes,  
\begin{equation}
    \h^{\ha}  = \sum_{\nu=3+d_r+1}^{3N}\qty(-\frac{1}{2}\nabla_{X_\nu}^2+\frac{1}{2}\omega_\nu^2X_\nu^2).
\end{equation}
The wavefunctions, $|\chi_{\nu,k}^\ha \rangle$, and energies, $\varepsilon_{\nu,k}^\ha$ of each simple harmonic oscillator is known analytically,
\begin{equation}\label{eq:ha_wf}
\langle X_\nu | \chi_{\nu,k}^\ha \rangle =  \frac{1}{\sqrt{2^kk!}}\qty(\frac{\omega_\nu}{\pi})^\frac{1}{4}\exp\qty[\frac{-\omega_\nu X_\nu^2}{2}]H_k\qty(\sqrt{\omega_\nu}X_\nu),
\end{equation}
\begin{equation}\label{eq:ha_en}
    \varepsilon_{\nu,k}^\ha = \qty(k+\frac{1}{2})\omega_\nu,
\end{equation}
with $H_k$ denoting the $k$-th order Hermite polynomial. Inserting the expression of $\varepsilon_{\nu,k}^\ha$ from Eq. \ref{eq:ha_en} into Eq. \ref{eq:partion_fn} yields, the partition function for the $\nu$-th harmonic oscillator,
\begin{equation}\label{eq:ha_pf}
   Q_{\nu}^\ha(T) = \sum_{k=0}^{\infty} \exp\qty[-\frac{\omega_\nu}{k_BT}(k+\frac{1}{2})] = \exp\qty(\frac{\omega_\nu}{2k_BT})n_B(\omega_\nu,T),
\end{equation}
where the Bose occupation factor is given by,
\begin{equation}\label{eq:be_occ}
    n_B(\omega,T) = \frac{1}{\exp(\omega/k_BT)-1}
\end{equation}
Because of the separation of Hamiltonian into $3N-3-d_r$ number of independent vibrational terms, the total vibrational wave function which is characterized by a vector of quantum numbers $\bfk = \qty(k_{3+d_r+1},k_{3+d_r+2},...,k_{3N})$ can be simplified by the product of the wave functions for each individual normal mode. The total partition function can be written analogously.
\begin{equation}\label{eq:ha_tot_wf_pf}
    \begin{aligned}
        |\chi_{\bfk}^\ha(R)\rangle = \prod_{\nu=3+d_r+1}^{3N}|\chi_{\nu,k_\nu}^\ha(R)\rangle, \\
        Q^\ha(T)=\prod_{\nu=3+d_r+1}^{3N}Q_\nu^\ha(T).
    \end{aligned}
\end{equation}
Utilizing the expressions from Eqs.\ref{eq:ha_wf}--\ref{eq:ha_tot_wf_pf}, Eq. \ref{eq:ensemble_avg} can be rewritten as,
\begin{equation}\label{eq:ha_avg}
    \langle \aobs \rangle_T^\ha = \int d\bfX W^\ha(\bfX,T) \aobs(\bfX) = \int d\bfX \qty[\prod_{\nu=3+d_r+1}^{3N} G(X_\nu;\sigma_{\nu,T})]\aobs(\bfX) 
\end{equation}
where the harmonic probability density, $W^\ha(\bfX,T)$, reduces to a product of independent Gaussian functions,
\begin{equation}\label{eq:gaussian}
    G(X_\nu;\sigma_{\nu,T}) = \frac{1}{\sqrt{2\pi\sigma_{\nu,T}^2}}\exp\qty(-\frac{X_{\nu}^2}{2\sigma_{\nu,T}^2}),
\end{equation}
with widths related to the Bose occupation factor:
\begin{equation}\label{eq:sigma}
    \sigma_{\nu,T}= \sqrt{\frac{2n_B\qty(\omega_\nu,T)+1}{2\omega_\nu}}.
\end{equation}

We note that though Eq. \ref{eq:ha_avg} is a result of harmonic approximation, it does not assume any explicit dependence of electronic observable $\aobs$ on nuclear coordinates ($\bfR$) or normal mode coordinates ($\bfX$). To further simplify the expression, we expand  $\aobs(\bfX)$ in Taylor series,
\begin{equation}\label{eq:obs_taylor}
    \aobs(\bfX) = \aobs(\bfzero) + \sum_{\nu}\eval{\pdv{\aobs}{X_\nu}}_{\bfzero}X_{\nu} + \sum_{\nu\nu'} \eval{\pdv{\aobs}{X_{\nu}}{X_{\nu'}}}_{\bfzero}X_{\nu}X_{\nu'} + ...
\end{equation}
and truncate the expansion after the second order. After inserting the resulting expression into Eq.\ref{eq:ha_avg}, we obtain the phonon renormalized electronic observable with quadratic (Q) approximation,
\begin{equation}\label{eq:qa_avg}
    \langle \aobs \rangle_T^\qa = \aobs(\bfzero) +  \sum_{\nu=3+d_r+1}^{3N}\frac{1}{2\omega_\nu}\eval{\pdv[2]{\aobs}{X_{\nu}}}_0\qty[n_B(\omega_{\nu},T)+\frac{1}{2}].
\end{equation}
We note that terms involving odd order of $X_\nu$ or cross-coupling terms such as $X_{\nu}X_{\nu'}$, with $\nu \neq \nu'$, do not appear because harmonic density is symmetric with respect to $\bfX = \bfzero$. For systems with strong anharmonicity, the vibrational density is no longer symmetric, and subsequently, both odd-order terms and cross-coupling terms become essential. 

\subsection{Stochastic Approach}
The stochastic approach employs Monte Carlo sampling of $W(\bfX,T)$ and subsequently utilizes Eq. \ref{eq:ha_avg} to compute phonon-renormalized electronic properties. In each Monte Carlo step, a displaced coordinate in normal mode is obtained, $\bfX^i=\bftau^{i}$, where, for $\nu>3+d_r$, the matrix elements, $\tau_{\nu}^i$, are a Gaussian distributed random number with zero mean and width $\sigma_{\nu,T}$, while the first $3+d_r$ matrix elements are set to zero. Afterward, $\bfX^i$'s are back-transformed to Cartesian coordinates, $\bfR^i$, using Eq.\ref{eq:norm2cart}, and subsequently, electronic observable $\aobs(\bfR^i)$ is computed. After $M$ Monte Carlo steps, Eq. \ref{eq:ha_avg} can be re-written as,
\begin{equation}\label{eq:mc_avg}
        \langle \aobs \rangle_T^{MC} = \frac{1}{M}\sum_{i=1}^{M}\aobs(\bfX^i) = \frac{1}{M}\sum_{i=1}^{M}\aobs(\bfR^i)
\end{equation}

Based on the Mean-value (MV) theorem and utilizing quadratic (Q) approximation, Monserrat proposed that there exists $2^{3N-3-d_r}$ mean-value positions, $\bfX_\text{MVQ}$, for which $\aobs(\bfX_\text{MVQ})\simeq\langle \aobs \rangle_T$,\cite{TL_Monserrat_PRB_2016} with $\bfX^i_\text{MVQ} =  \bfs^i\boldsymbol{\sigma}_T$, where the matrix elements of $\boldsymbol{\sigma}_T$ is given by Eq. \ref{eq:sigma}, and $\bfs^i$ is a matrix with first $3+d_r$ elements set to zero and remaining $3N-3-d_r$ elements are either +1 or -1. It was shown that a Monte Carlo algorithm that samples random signs, $s^i$, has a faster convergence of $\langle \aobs \rangle_T^{MC}$ than the one that samples  random numbers, $\boldsymbol{\tau}^i$, from a Gaussian distribution. \cite{TL_Monserrat_PRB_2016} Later, Zacharius and Giustino proposed a one-shot (OS) algorithm, \cite{Zacharias_PRB_2016} in which the signs are chosen according to,
\begin{equation}
\begin{split}
    s_{\nu} &= (-1)^{\nu-4-d_r} \text{ for } \nu > 3+d_r \\
    &= 0, \text{ for } \nu \le 3+d_r
\end{split}
\end{equation}
and they showed that only a single first-principle calculation on the obtained atomic configuration, $\bfX=\bfs\boldsymbol{\sigma}_T$ is sufficient to converge of $\langle \aobs \rangle_T$. They also proposed that an additional first-principles calculation on the antithetic pair of the chosen atomic configuration, i.e., $\bfX=-\bfs\boldsymbol{\sigma}_T$, can improve the result, which we adopted here. 
\begin{equation}\label{eq:mc_avg}
        \langle \aobs \rangle_T^{OS} = \frac{1}{2}\qty[\aobs\qty(\bfs\boldsymbol{\sigma}_T) + \aobs\qty(-\bfs\boldsymbol{\sigma}_T)]
\end{equation}

\subsection{Frozen Phonon Approach}
A frozen phonon (FP) approach utilizes Eq. \ref{eq:qa_avg} to compute phonon-renormalized electronic properties. To compute the first and second derivatives of the observable, $\aobs'$, $\aobs''$, respectively, we displaced the nuclei to $X_{\nu}=+h_\nu$ and $X_{\nu}=-h_\nu$ along each phonon coordinates, and subsequently, used the central difference formula.
\begin{equation} \label{eq:central_diff1}
\aobs'(X_{\nu}=0) = \eval{\pdv{\aobs(X_{\nu})}{X_{\nu}}}_{X_{\nu}=0}= \frac{\aobs(+ h_\nu)-\aobs(-h_\nu)}{2h_\nu}
\end{equation}
\begin{equation} \label{eq:central_diff2}
    \aobs''(X_{\nu}=0) = \eval{\pdv[2]{\aobs(X_{\nu})}{X_{\nu}}}_{X_{\nu}=0}=  \frac{\aobs(+ h_\nu)+\aobs(-h_\nu)-2\aobs(0)}{h_{\nu}^2},
\end{equation}
We chose the displacements, $h_\nu$, such that the potential energy difference, $\delta\V^{H}$, is the same for all modes, assuming a parabolic dependence of potential energy on each normal mode, i.e., 
\begin{equation}\label{eq:enmfd}
    h_\nu = \frac{\sqrt{2\delta\V^{H}}}{\omega_\nu}.
\end{equation}

Throughout this work, our electronic observable $(\aobs)$ would be either valence and conduction band (VB) energies, $E_n$, with $n$ denoting the band index, or the band gap, $E_g$. The second derivative of $n$-th band energies with respect to $\nu$-th phonon mode scaled by the phonon frequency, which appears in the Eq. \ref{eq:qa_avg} when $\aobs=E_n$, is called the electron-phonon coupling energy (EPCE) of band $n$ with respect to phonon mode $\nu$. 
\begin{equation}\label{eq:epce}
    \text{EPCE}_{n,\nu} = \frac{1}{2\omega_\nu}\eval{\pdv[2]{E_n}{X_{\nu}}}_0
\end{equation}
It is evident from Eq. \ref{eq:qa_avg}, the electron-phonon renormalization of $n$-th band, 
\begin{equation}
    \Delta E_n(T) = \langle E_n \rangle_T - E_n(\bfzero)
\end{equation}
reduces to,
\begin{equation}
     \Delta E_n^Q(0) = \sum_{\nu=3+d_r+1}^{N}\text{EPCE}_{n,\nu}
\end{equation}
within quadratic approximation at 0 K. EPCEs can be computed using the FP method and it describes the contributions of each mode towards the total electron-phonon renormalization of band energies.

Depending on the displacement chosen for each phonon mode, nearby eigenstates may cross each other, as we shall see during the discussion. In such cases, while calculating $\eval{\pdv{E_n}{X_{\nu}}}_0$ and $\eval{\pdv[2]{E_n}{X_{\nu}}}_0$ by applying Eq. \ref{eq:central_diff1} and \ref{eq:central_diff2}, respectively, it is necessary to ensure that the electronic eigenvalues used for the positive and negative displacements correspond to the same eigenstate. Following earlier works \cite{FPH_Bester_2013,FPH_Bester_2019}, we ensured the correct correspondence by projecting wavefunctions with negative displacements on wavefunctions with positive displacements.

\clearpage
\section{Additional Figures}
\subsection{Mode Resolved EPCEs and Anharmonic Measures }\label{sec:epce}

\begin{figure}[tbhp]
\centering
\includegraphics[width=10cm]{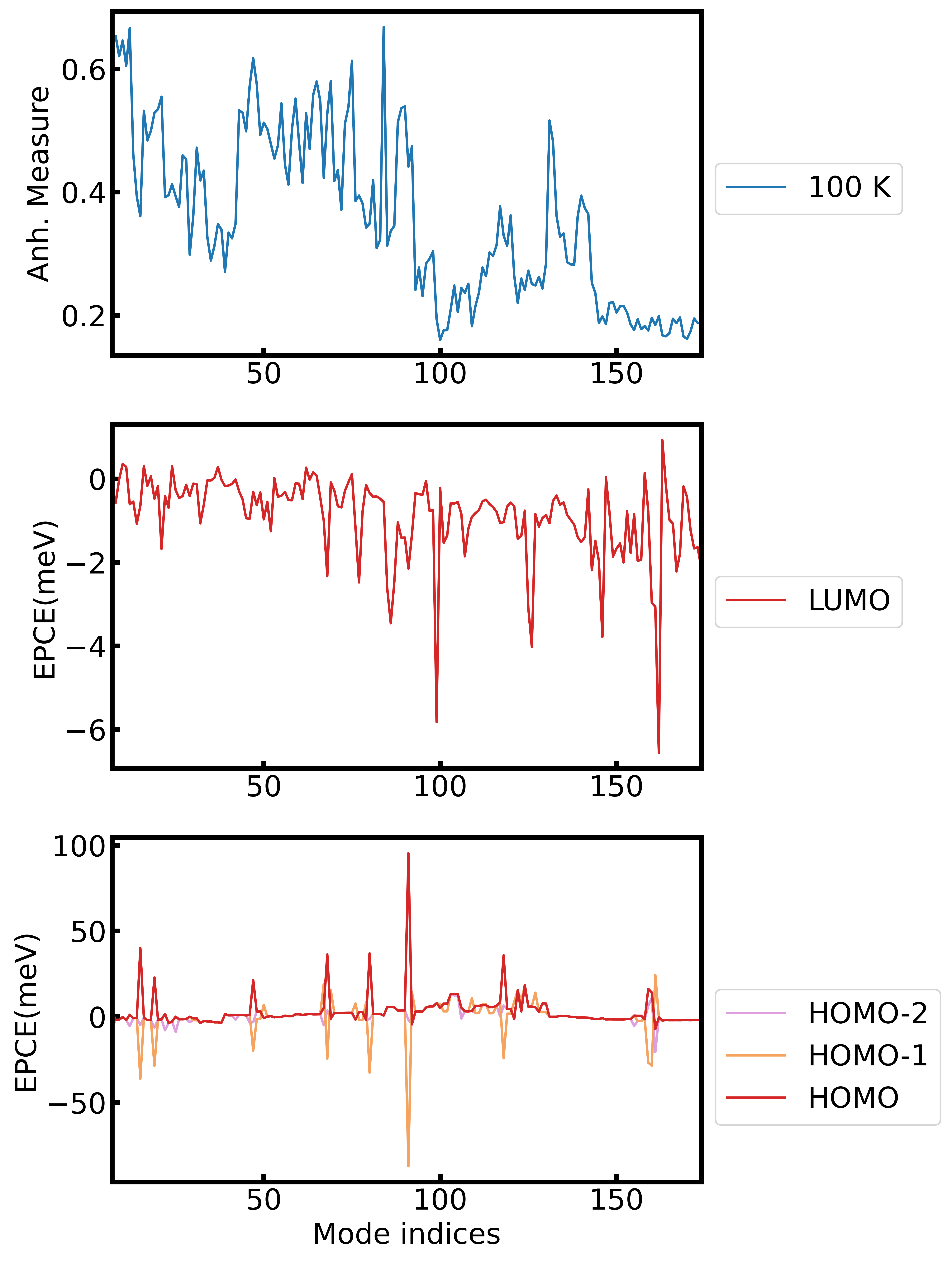}
\caption{ Mode-resolved anharmonic measures and electron-phonon coupling energies (EPCEs) of the lowest and highest occupied molecular orbitals for an isolated pentamantane molecule as obtained with the SCAN functional. \textbf{Top panel:} Mode-resolved anharmonic measures computed from the trajectory obtained with the quantum thermostatted molecular dynamics simulations at 100 K. \textbf{Middle panel:} Mode-resolved EPCEs of the lowest unoccupied molecular orbital (LUMO) computed using the frozen phonon approach. \textbf{Bottom panel:} Mode-resolved EPCEs of the highest occupied molecular orbital (HOMO), HOMO-1, and HOMO-2 levels computed using the frozen phonon approach.    
}
\label{fig:epce-penta-mol}
\end{figure}

\begin{figure}[tbhp]
\centering
\includegraphics[width=10cm]{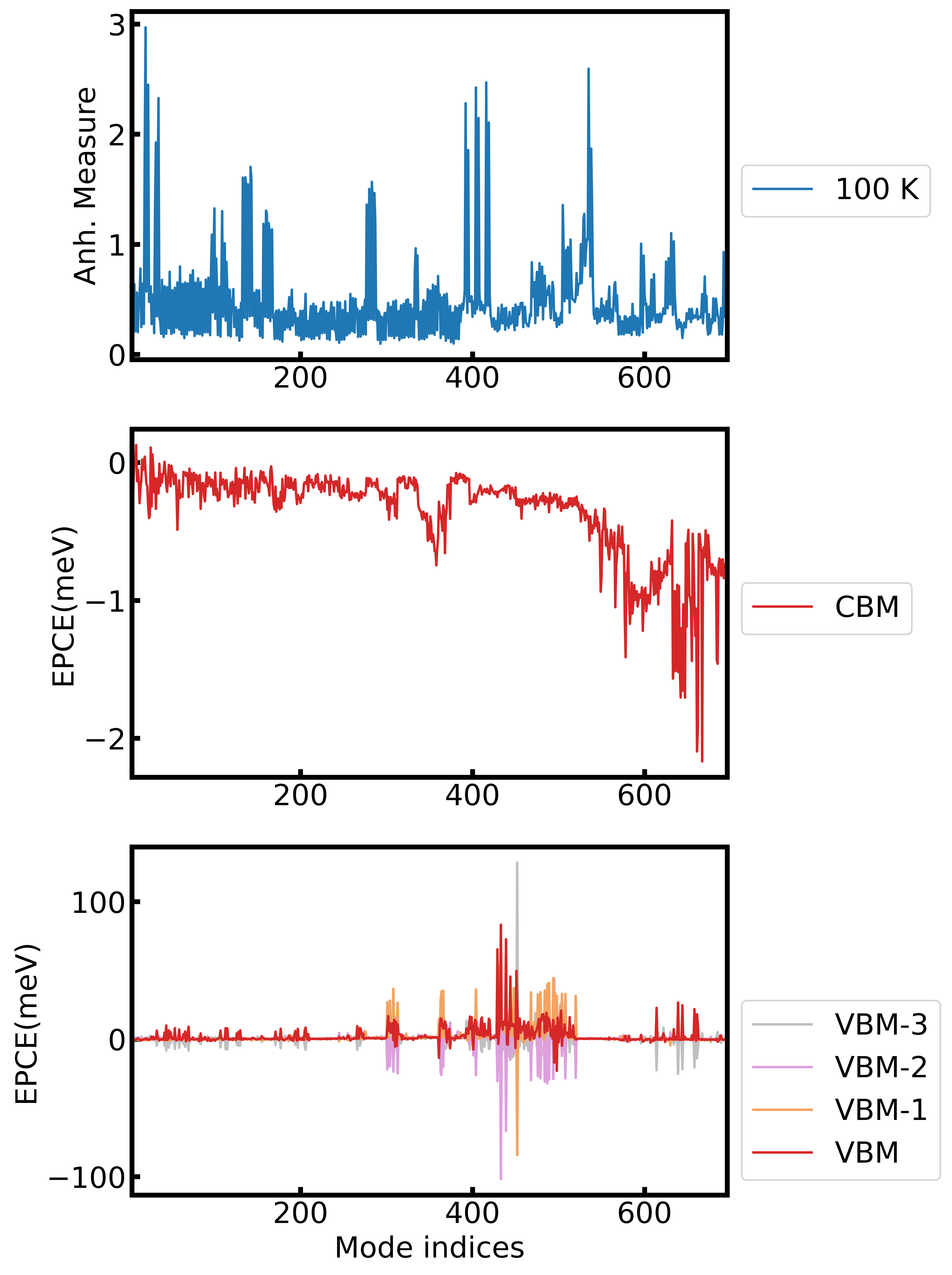}
\caption{ Mode-resolved anharmonic measures and electron-phonon coupling energies (EPCEs) of the valence and conduction bands for the pentamantane molecular crystal as obtained with the SCAN functional. \textbf{Top panel:} Mode-resolved anharmonic measures computed from the trajectory obtained with the quantum thermostatted molecular dynamics simulations at 100 K. \textbf{Middle panel:} Mode-resolved EPCEs of the conduction band minimum (CBM) computed using a frozen phonon calculation. \textbf{Bottom panel:} Mode-resolved EPCEs of the valence band maximum (VBM), VBM-1, VBM-2 and VBM-3 levels computed using the frozen phonon approach.    
}
\label{fig:epce-penta-cry}
\end{figure}

\begin{figure}[tbhp]
\centering
\includegraphics[width=10cm]{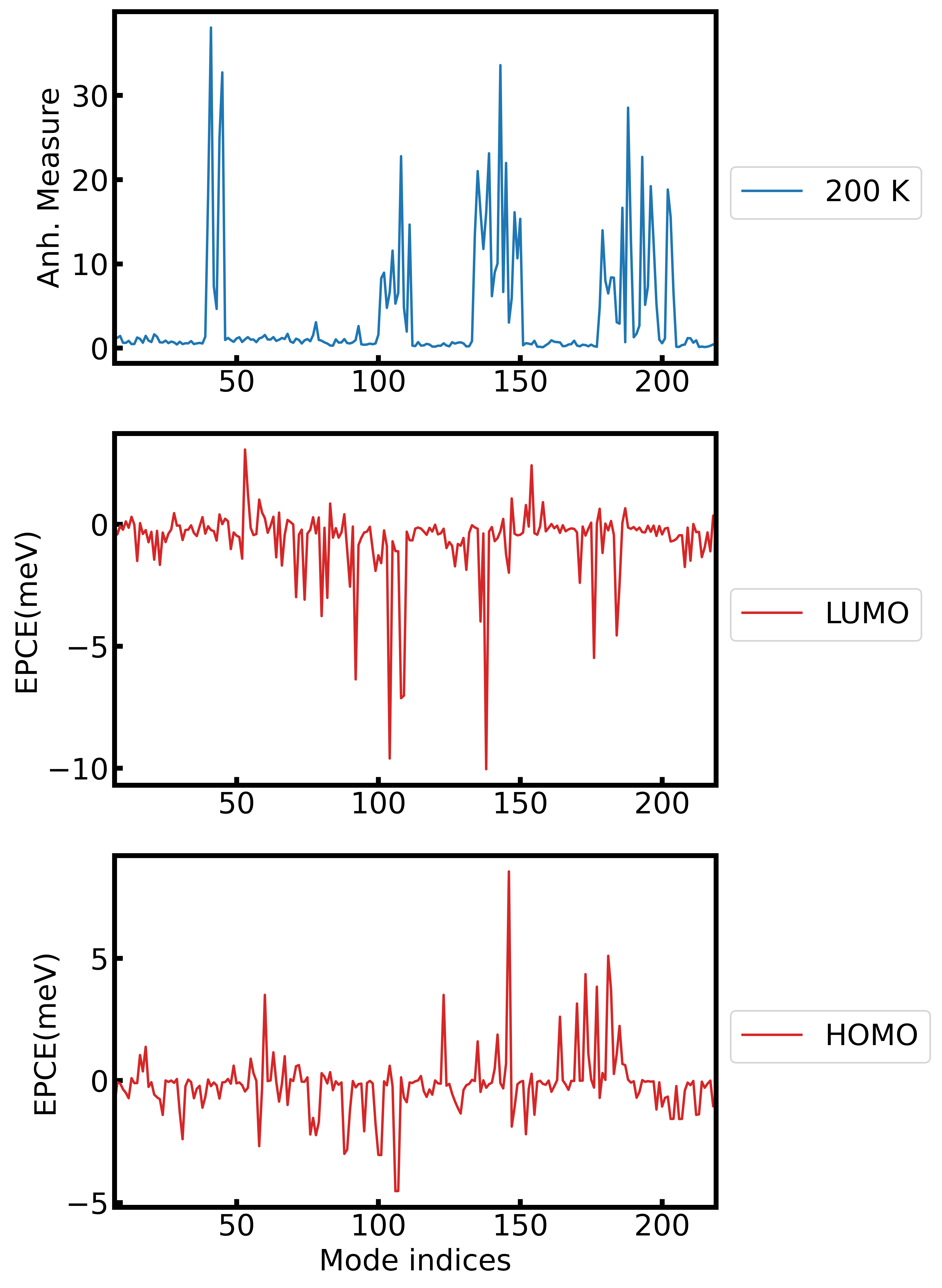}
\caption{Mode-resolved anharmonic measures and electron-phonon coupling energies (EPCEs) of the lowest and highest occupied molecular orbitals for an isolated NAI-DMAC molecule as obtained with the PBE functional. \textbf{Top panel:} Mode-resolved anharmonic measures computed from the trajectory obtained with the quantum thermostatted molecular dynamics simulations at 200 K. \textbf{Middle panel:} Mode-resolved EPCEs of the lowest unoccupied molecular orbital (LUMO) computed using the frozen phonon approach. \textbf{Bottom panel:} Mode-resolved EPCEs of the highest occupied molecular orbital (HOMO) computed using the frozen phonon approach. 
}
\label{fig:epce-naidmac-mol}
\end{figure}

\begin{figure}[tbhp]
\centering
\includegraphics[width=10cm]{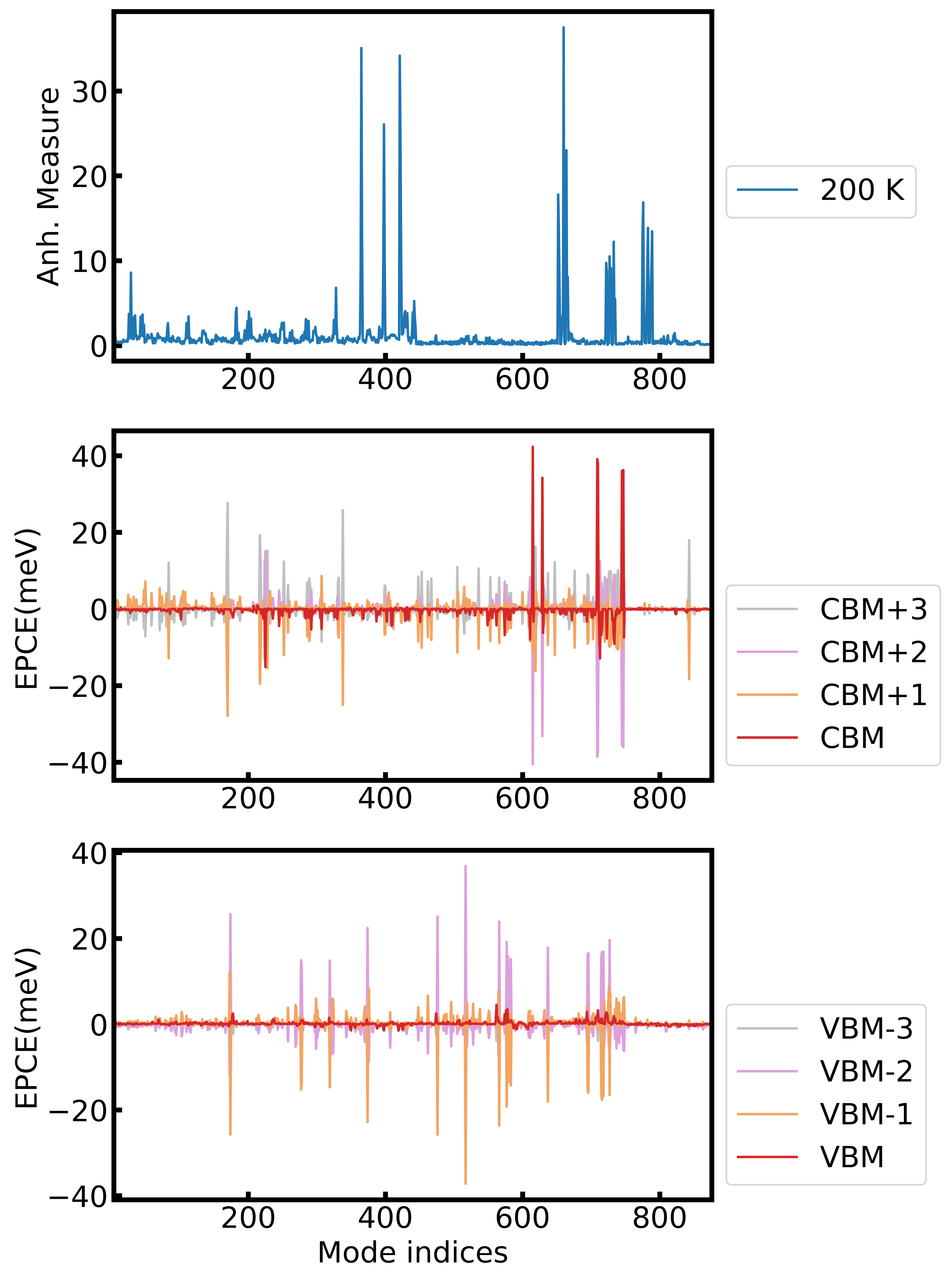}
\caption{ Mode-resolved anharmonic measures and electron-phonon coupling energies (EPCEs) of the valence and conduction bands for NAI-DMAC molecular crystal as obtained with the PBE functional. \textbf{Top panel:} Mode-resolved anharmonic measures computed from the trajectory obtained with the quantum thermostatted molecular dynamics simulations at 200 K. \textbf{Middle panel:} Mode-resolved EPCEs of the conduction band minimum (CBM), CBM+1, CBM+2, and CBM+3 levels computed using the frozen phonon approach. \textbf{Bottom panel:} Mode-resolved EPCEs of the valence band maximum (VBM), VBM-1, VBM-2 and VBM-3 levels computed using the frozen phonon approach.    
}
\label{fig:epce-naidmac-cry}
\end{figure}

\clearpage
\subsection{Effect of torsion on the potential energy surface and the band gap of an isolated NAI-DMAC molecule}\label{sec:dihedral}

\begin{figure}[tbhp]
\centering
\includegraphics[width=12cm]{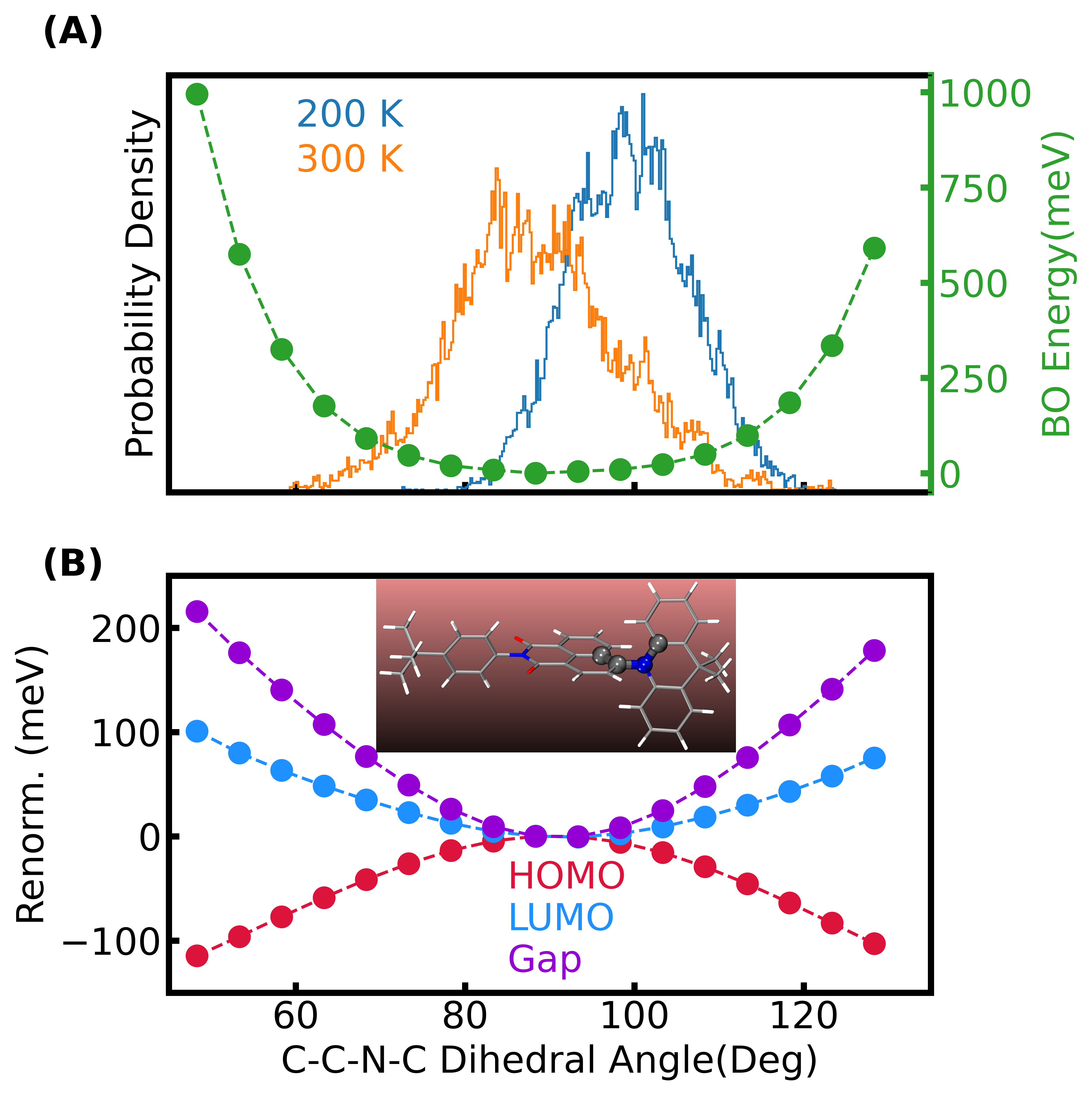}
\caption{ Panel A: Probability distribution of the C-C-N-C dihedral angle (highlighted as spheres in the inset of panel B) for the isolated molecule at different temperatures as obtained from QTMD simulations. The green circles represent the Born-Oppenheimer energy when PES is scanned by varying the dihedral angle for the isolated molecule. Panel B: The HOMO, LUMO, and fundamental renormalizations  for the isolated molecule when only the dihedral angle is varied. See also Figs. 5,6 and 9 of ref. \cite{Francese_PCCP_2022} and the discussions therein.    
}
\label{fig:nai-dmac-dihed-scan}
\end{figure}

\begin{figure}[tbhp]
\centering
\includegraphics[width=16cm]{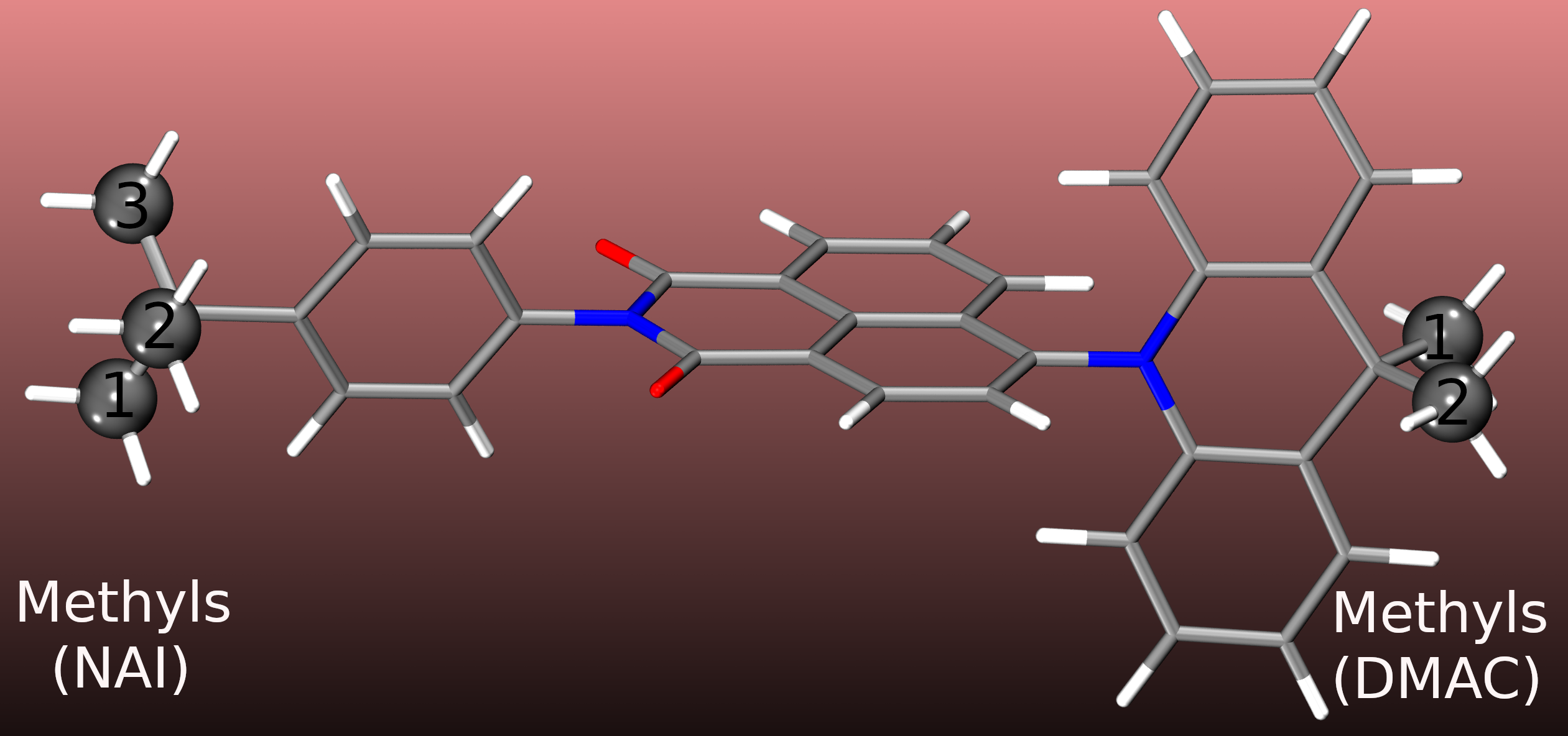}
\caption{ Three and two methyl groups of the NAI and DMAC units, respectively, of the isolated NAIDMAC molecules that are rotated relative to the geometry optimized positions. The central carbon atoms of these methyl groups are highlighted as gray spheres.     
}
\label{fig:methyl-labels}
\end{figure}

\begin{figure}[tbhp]
\centering
\includegraphics[width=12cm]{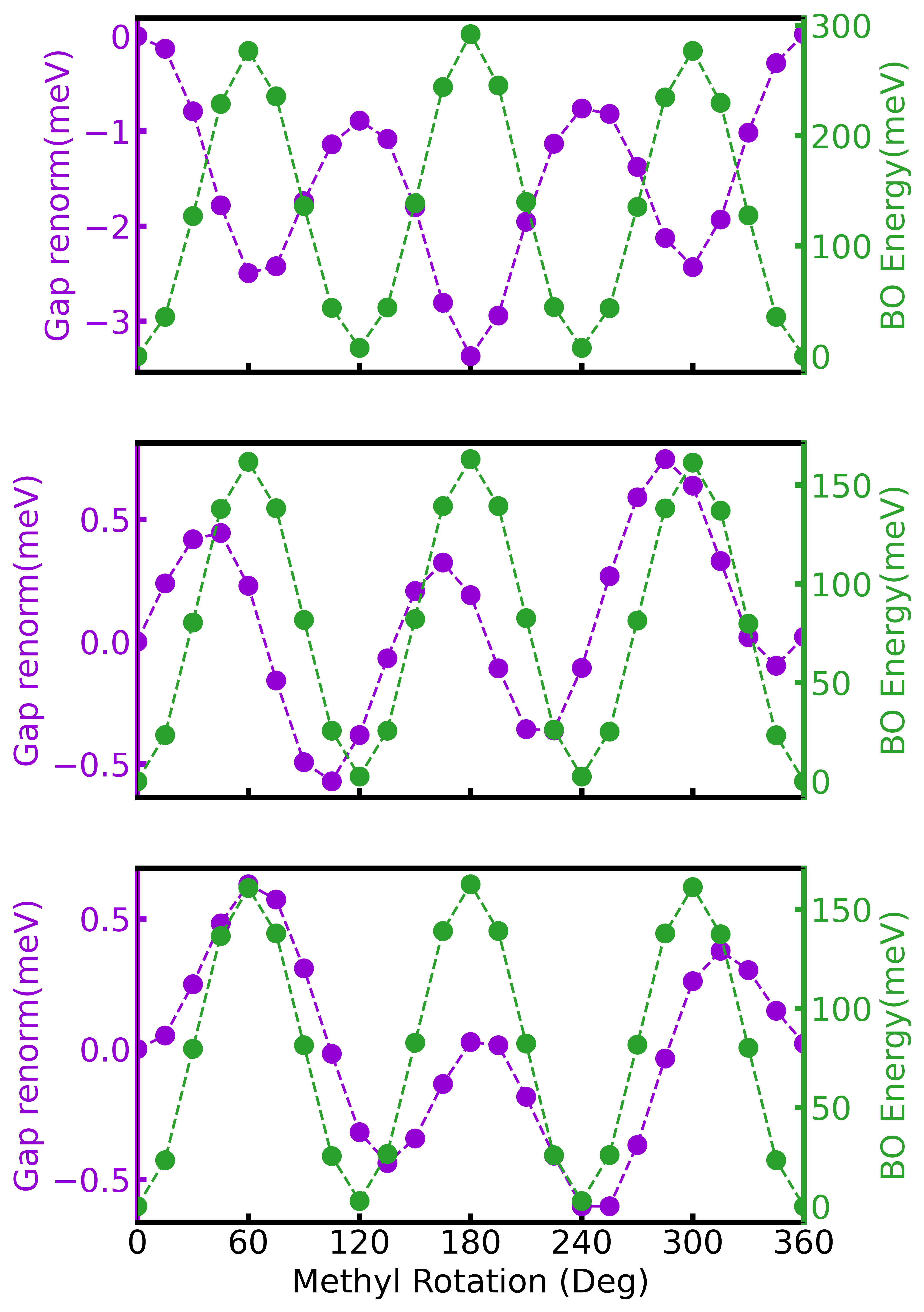}
\caption{Born-Oppenheimer(BO) energies and the HOMO-LUMO gap renormalizations obtained with the PBE functional for an isolated NAI-DMAC molecule when its three methyl groups within the NAI unit are rotated relative to the geometry optimized position. The top, middle, and bottom panels show the results for the NAI methyl groups labeled with numbers 1, 2, and 3, respectively, in Fig. \ref{fig:methyl-labels}.}
\label{fig:nai-methyl}
\end{figure}

\begin{figure}[tbhp]
\centering
\includegraphics[width=12cm]{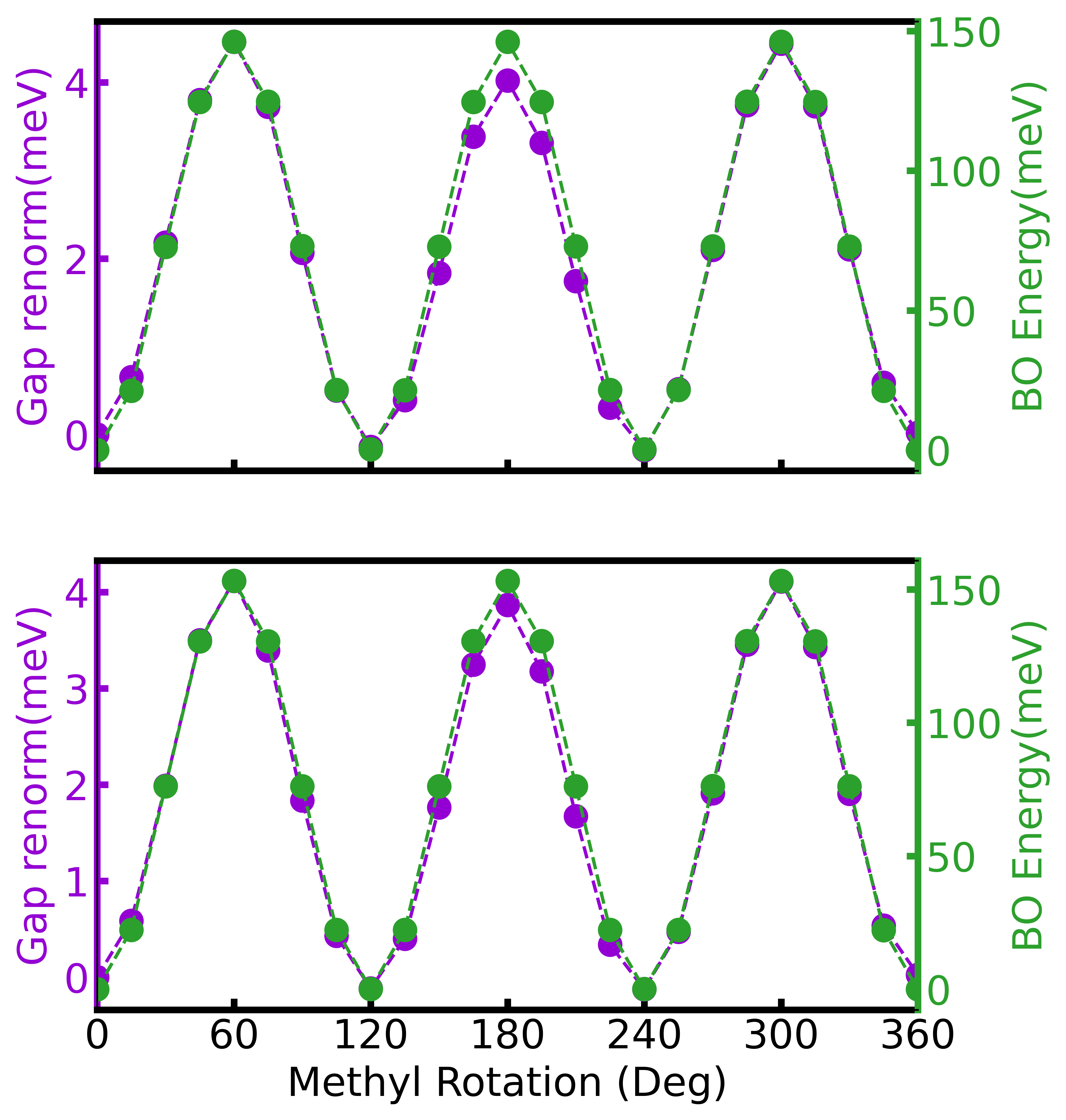}
\caption{Born-Oppenheimer(BO) energies and the HOMO-LUMO gap renormalizations obtained with the PBE functional for an isolated NAI-DMAC molecule when its two methyl groups within the DMAC unit are rotated relative to the geometry optimized position. The top and bottom panels show the results for the DMAC methyl groups labeled with numbers 1 and 2, respectively, in Fig. \ref{fig:methyl-labels}.}
\label{fig:dmac-methyl}
\end{figure}

\clearpage
%\bibliography{references}
\providecommand{\latin}[1]{#1}
\makeatletter
\providecommand{\doi}
  {\begingroup\let\do\@makeother\dospecials
  \catcode`\{=1 \catcode`\}=2 \doi@aux}
\providecommand{\doi@aux}[1]{\endgroup\texttt{#1}}
\makeatother
\providecommand*\mcitethebibliography{\thebibliography}
\csname @ifundefined\endcsname{endmcitethebibliography}
  {\let\endmcitethebibliography\endthebibliography}{}